%% Draft Mode
%\def\mydraft{label}           %turn on label
%\def\mydraft{no label}        %turn off label
%\def\drafttitle{minimal1.tex}   %display title on top-right

%%%%%%%%%%%%%%%%%%%%%%%%%%%%%%%%
%                                                                                                              %
% Vasilis Niarchos (Niels Bohr Institute)                                          %
%                                                                                                              %
%%%%%%%%%%%%%%%%%%%%%%%%%%%%%%%%

\input lanlmac

\input epsf

%Macro for figure
\newcount\figno
\figno=0
\def\fig#1#2#3{
\par\begingroup\parindent=0pt\leftskip=1cm\rightskip=1cm\parindent=0pt
\baselineskip=11pt
\global\advance\figno by 1
\midinsert
\epsfxsize=#3
\centerline{\epsfbox{#2}}
\vskip 12pt
\centerline{{\bf Fig. \the\figno:~~} #1}\par
\endinsert\endgroup\par
}
\def\figlabel#1{\xdef#1{\the\figno}}

%%%%%%%
% Older defs

\def\journal#1&#2(#3){\unskip, \sl #1\ \bf #2 \rm(19#3) }
\def\andjournal#1&#2(#3){\sl #1~\bf #2 \rm (19#3) }

\def\frac#1#2{{#1\over#2}}

\def\d{\partial}

\def\inbar{\,\vrule height1.5ex width.4pt depth0pt}
\def\IC{\relax\hbox{$\inbar\kern-.3em{\rm C}$}}
\def\IR{\relax{\rm I\kern-.18em R}}
\def\IP{\relax{\rm I\kern-.18em P}}
\def\IZ{\relax{\rm I\kern-.18em Z}}

%
%%%%%%%%%%%%%%%%%%%%%%%%%%%%%%%%%%%%
%

%
\catcode`\@=11
\def\slash#1{\mathord{\mathpalette\c@ncel{#1}}}
\overfullrule=0pt

\def\CC{{\cal C}}
\def\DD{{\cal D}}

\def\FF{{\cal F}}

\def\II{{\cal I}}

\def\NN{{\cal N}}
\def\OO{{\cal O}}

\def\TT{{\cal T}}

\def\VV{{\cal V}}

\def\ZZ{{\cal Z}}

\def\underrel#1\over#2{\mathrel{\mathop{\kern\z@#1}\limits_{#2}}}

\catcode`\@=12

%%%%%%%%%%%%%%%%%%%%%%%%%%%%%%%%%%%%%%%%%%%%%%%%%

%

%%%%%%%%%%%%%%%%%%%%%%%%%%%%%%%%%%%%%%%%%%%%%%%%%
% new defs:

% Something to deal with sub-sub-sections

\def\unlockat{\catcode`\@=11}
\def\lockat{\catcode`\@=12}

\unlockat

%%% Something to deal with sub-sub-sections

\def\newsec#1{\global\advance\secno by1\message{(\the\secno. #1)}
\global\subsecno=0\global\subsubsecno=0\eqnres@t\noindent
{\bf\the\secno. #1}
\writetoca{{\secsym} {#1}}\par\nobreak\medskip\nobreak}
\global\newcount\subsecno \global\subsecno=0
\def\subsec#1{\global\advance\subsecno
by1\message{(\secsym\the\subsecno. #1)}
\ifnum\lastpenalty>9000\else\bigbreak\fi\global\subsubsecno=0
\noindent{\it\secsym\the\subsecno. #1}
\writetoca{\string\quad {\secsym\the\subsecno.} {#1}}
\par\nobreak\medskip\nobreak}
\global\newcount\subsubsecno \global\subsubsecno=0
\def\subsubsec#1{\global\advance\subsubsecno by1
\message{(\secsym\the\subsecno.\the\subsubsecno. #1)}
\ifnum\lastpenalty>9000\else\bigbreak\fi
\noindent\quad{\secsym\the\subsecno.\the\subsubsecno.}{#1}
\writetoca{\string\qquad{\secsym\the\subsecno.\the\subsubsecno.}{#1}}
\par\nobreak\medskip\nobreak}

\def\subsubseclab#1{\DefWarn#1\xdef
#1{\noexpand\hyperref{}{subsubsection}%
{\secsym\the\subsecno.\the\subsubsecno}%
{\secsym\the\subsecno.\the\subsubsecno}}%
\writedef{#1\leftbracket#1}\wrlabeL{#1=#1}}% Macros for boxes
\lockat

% End of Older Defs
%%%%%%%%%%%%%

%\def\tilde{\widetilde}
\newcount\figno
\figno=0
\def\fig#1#2#3{
\par\begingroup\parindent=0pt\leftskip=1cm\rightskip=1cm\parindent=0pt
\baselineskip=11pt
\global\advance\figno by 1
\midinsert
\epsfxsize=#3
\centerline{\epsfbox{#2}}
\vskip 12pt
{\bf Fig.\ \the\figno: } #1\par
\endinsert\endgroup\par
}
\def\figlabel#1{\xdef#1{\the\figno}}
\def\encadremath#1{\vbox{\hrule\hbox{\vrule\kern8pt\vbox{\kern8pt
\hbox{$\displaystyle #1$}\kern8pt}
\kern8pt\vrule}\hrule}}
%
%

%%% Paragraphs

%%% special math symbols
\font\cmss=cmss10
\font\cmsss=cmss10 at 7pt
\def\rlx{\relax\leavevmode}
\def\inbar{\vrule height1.5ex width.4pt depth0pt}
\def\IC{\relax\,\hbox{$\inbar\kern-.3em{\rm C}$}}
\def\IN{\relax{\rm I\kern-.18em N}}
\def\IP{\relax{\rm I\kern-.18em P}}
\def\ZZ{\rlx\leavevmode\ifmmode\mathchoice{\hbox{\cmss Z\kern-.4em Z}}
 {\hbox{\cmss Z\kern-.4em Z}}{\lower.9pt\hbox{\cmsss Z\kern-.36em Z}}
 {\lower1.2pt\hbox{\cmsss Z\kern-.36em Z}}\else{\cmss Z\kern-.4em
 Z}\fi}
%%% misc.
\def\IZ{\relax\ifmmode\mathchoice
{\hbox{\cmss Z\kern-.4em Z}}{\hbox{\cmss Z\kern-.4em Z}}
{\lower.9pt\hbox{\cmsss Z\kern-.4em Z}}
{\lower1.2pt\hbox{\cmsss Z\kern-.4em Z}}\else{\cmss Z\kern-.4em
Z}\fi}
%%% misc.
\def\IZ{\relax\ifmmode\mathchoice
{\hbox{\cmss Z\kern-.4em Z}}{\hbox{\cmss Z\kern-.4em Z}}
{\lower.9pt\hbox{\cmsss Z\kern-.4em Z}}
{\lower1.2pt\hbox{\cmsss Z\kern-.4em Z}}\else{\cmss Z\kern-.4em
Z}\fi}

\def\narrowplus{\kern -.04truein + \kern -.03truein}
\def\narrowminus{- \kern -.04truein}
\def\narrowminussub{\kern -.02truein - \kern -.01truein}

\def\IZ{\relax\ifmmode\mathchoice
{\hbox{\cmss Z\kern-.4em Z}}{\hbox{\cmss Z\kern-.4em Z}}
{\lower.9pt\hbox{\cmsss Z\kern-.4em Z}}
{\lower1.2pt\hbox{\cmsss Z\kern-.4em Z}}\else{\cmss Z\kern-.4em
Z}\fi}
\def\IB{\relax{\rm I\kern-.18em B}}
\def\IC{{\relax\hbox{$\inbar\kern-.3em{\rm C}$}}}
\def\ID{\relax{\rm I\kern-.18em D}}
\def\IE{\relax{\rm I\kern-.18em E}}
\def\IF{\relax{\rm I\kern-.18em F}}
\def\IG{\relax\hbox{$\inbar\kern-.3em{\rm G}$}}
\def\IGa{\relax\hbox{${\rm I}\kern-.18em\Gamma$}}
\def\IH{\relax{\rm I\kern-.18em H}}
\def\II{\relax{\rm I\kern-.18em I}}
\def\IK{\relax{\rm I\kern-.18em K}}
\def\IP{\relax{\rm I\kern-.18em P}}
%\def\IX{\relax{\rm X\kern-.01em X}}
%this doesn't work

\font\cmss=cmss10 \font\cmsss=cmss10 at 7pt
\def\IR{\relax{\rm I\kern-.18em R}}

%\def\mp{M_{\rm p}}

%

%
%       \eqn\label{a+b=c}       gives displayed equation, numbered
%                               consecutively within sections.
%     \eqnn and \eqna define labels in advance (of eqalign?)
%
\def\eqnn#1{\xdef #1{(\secsym\the\meqno)}\writedef{#1\leftbracket#1}%
\global\advance\meqno by1\wrlabeL#1}
\def\eqna#1{\xdef #1##1{\hbox{$(\secsym\the\meqno##1)$}}
\writedef{#1\numbersign1\leftbracket#1{\numbersign1}}%
\global\advance\meqno by1\wrlabeL{#1$\{\}$}}
\def\eqn#1#2{\xdef #1{(\secsym\the\meqno)}\writedef{#1\leftbracket#1}%
\global\advance\meqno by1$$#2\eqno#1\eqlabeL#1$$}

%%%%%%%%%%%%%%%%%%%%
% References

%\MukhiZB
\lref\MukhiZB{
S.~Mukhi and C.~Vafa,
``Two-dimensional black hole as a topological coset model of c = 1 string
theory,''
Nucl.\ Phys.\ B {\bf 407}, 667 (1993)
[arXiv:hep-th/9301083].
%%CITATION = HEP-TH 9301083;%%
}

%\AharonyVK
\lref\AharonyVK{
O.~Aharony, B.~Fiol, D.~Kutasov and D.~A.~Sahakyan,
``Little string theory and heterotic/type II duality,''
Nucl.\ Phys.\ B {\bf 679}, 3 (2004)
[arXiv:hep-th/0310197].
%%CITATION = HEP-TH 0310197;%%
}

%\RibaultWP
\lref\RibaultWP{
  S.~Ribault and J.~Teschner,
  ``H(3)+ correlators from Liouville theory,''
  arXiv:hep-th/0502048.
  %%CITATION = HEP-TH 0502048;%%
}

%\RibaultMS
\lref\RibaultMS{
  S.~Ribault,
  ``Knizhnik-Zamolodchikov equations and spectral flow in AdS3 string theory,''
  arXiv:hep-th/0507114.
  %%CITATION = HEP-TH 0507114;%%
}

%\GiribetIX
\lref\GiribetIX{
  G.~Giribet and Y.~Nakayama,
  ``The Stoyanovsky-Ribault-Teschner map and string scattering amplitudes,''
  arXiv:hep-th/0505203.
  %%CITATION = HEP-TH 0505203;%%
}

%\AganagicQJ
\lref\AganagicQJ{
  M.~Aganagic, R.~Dijkgraaf, A.~Klemm, M.~Marino and C.~Vafa,
  ``Topological strings and integrable hierarchies,''
  arXiv:hep-th/0312085.
  %%CITATION = HEP-TH 0312085;%%
}

%\GhoshalWM
\lref\GhoshalWM{
  D.~Ghoshal and C.~Vafa,
  ``C = 1 string as the topological theory of the conifold,''
  Nucl.\ Phys.\ B {\bf 453}, 121 (1995)
  [arXiv:hep-th/9506122].
  %%CITATION = HEP-TH 9506122;%%
}

%\OoguriWJ
\lref\OoguriWJ{
  H.~Ooguri and C.~Vafa,
  ``Two-Dimensional Black Hole and Singularities of CY Manifolds,''
  Nucl.\ Phys.\ B {\bf 463}, 55 (1996)
  [arXiv:hep-th/9511164].
  %%CITATION = HEP-TH 9511164;%%
}

%\StoyanovskyPG
\lref\StoyanovskyPG{
  A.~V.~Stoyanovsky,
  ``A relation between the Knizhnik--Zamolodchikov and
  Belavin--Polyakov--Zamolodchikov systems of partial differential equations,''
  arXiv:math-ph/0012013.
  %%CITATION = MATH-PH 0012013;%%
}

%\AharonyXN
\lref\AharonyXN{
  O.~Aharony, A.~Giveon and D.~Kutasov,
  ``LSZ in LST,''
  Nucl.\ Phys.\ B {\bf 691}, 3 (2004)
  [arXiv:hep-th/0404016].
  %%CITATION = HEP-TH 0404016;%%
}

%\DijkgraafDJ
\lref\DijkgraafDJ{
  R.~Dijkgraaf, H.~L.~Verlinde and E.~P.~Verlinde,
  ``Topological Strings In $D < 1$,''
  Nucl.\ Phys.\ B {\bf 352}, 59 (1991).
  %%CITATION = NUPHA,B352,59;%%
}

%\WarnerZH
\lref\WarnerZH{
  N.~P.~Warner,
  ``N=2 supersymmetric integrable models and topological field theories,''
  arXiv:hep-th/9301088.
  %%CITATION = HEP-TH 9301088;%%
}

%\WittenZZ
\lref\WittenZZ{
  E.~Witten,
  ``Mirror manifolds and topological field theory,''
  arXiv:hep-th/9112056.
  %%CITATION = HEP-TH 9112056;%%
}

%\OhtaEH
\lref\OhtaEH{
  N.~Ohta and H.~Suzuki,
  ``Bosonization of a topological coset model and noncritical string theory,''
  Mod.\ Phys.\ Lett.\ A {\bf 9}, 541 (1994)
  [arXiv:hep-th/9310180].
  %%CITATION = HEP-TH 9310180;%%
}

%\LercheUY
\lref\LercheUY{
  W.~Lerche, C.~Vafa and N.~P.~Warner,
  ``Chiral Rings In N=2 Superconformal Theories,''
  Nucl.\ Phys.\ B {\bf 324}, 427 (1989).
  %%CITATION = NUPHA,B324,427;%%
}

%\GiveonZM
\lref\GiveonZM{
  A.~Giveon, D.~Kutasov and O.~Pelc,
  ``Holography for non-critical superstrings,''
  JHEP {\bf 9910}, 035 (1999)
  [arXiv:hep-th/9907178].
  %%CITATION = HEP-TH 9907178;%%
}

%\OoguriFP
\lref\OoguriFP{
  H.~Ooguri and C.~Vafa,
  ``Geometry of N=2 strings,''
  Nucl.\ Phys.\ B {\bf 361}, 469 (1991).
  %%CITATION = NUPHA,B361,469;%%
}

%\OoguriIE
\lref\OoguriIE{
  H.~Ooguri and C.~Vafa,
  ``N=2 heterotic strings,''
  Nucl.\ Phys.\ B {\bf 367}, 83 (1991).
  %%CITATION = NUPHA,B367,83;%%
}

%\AharonyVK
\lref\AharonyVK{
  O.~Aharony, B.~Fiol, D.~Kutasov and D.~A.~Sahakyan,
  ``Little string theory and heterotic/type II duality,''
  Nucl.\ Phys.\ B {\bf 679}, 3 (2004)
  [arXiv:hep-th/0310197].
  %%CITATION = HEP-TH 0310197;%%
}

%\KonechnyDF
\lref\KonechnyDF{
  A.~Konechny, A.~Parnachev and D.~A.~Sahakyan,
  ``The ground ring of N = 2 minimal string theory,''
  arXiv:hep-th/0507002.
  %%CITATION = HEP-TH 0507002;%%
}

%\TakayanagiYJ
\lref\TakayanagiYJ{
  T.~Takayanagi,
  ``Notes on S-matrix of non-critical N = 2 string,''
  arXiv:hep-th/0507065.
  %%CITATION = HEP-TH 0507065;%%
}

%\KutasovUF
\lref\KutasovUF{
  D.~Kutasov,
  ``Introduction to little string theory,''
%\href{http://www.slac.stanford.edu/spires/find/hep/www?irn=5015588}{SPIRES entry}
{\it Prepared for ICTP Spring School on Superstrings and Related Matters, Trieste, Italy, 2-10 Apr 2001}
}

%\AharonyKS
\lref\AharonyKS{
  O.~Aharony,
  ``A brief review of 'little string theories',''
  Class.\ Quant.\ Grav.\  {\bf 17}, 929 (2000)
  [arXiv:hep-th/9911147].
  %%CITATION = HEP-TH 9911147;%%
}

%\DijkgraafFC
\lref\DijkgraafFC{
  R.~Dijkgraaf and C.~Vafa,
  ``Matrix models, topological strings, and supersymmetric gauge theories,''
  Nucl.\ Phys.\ B {\bf 644}, 3 (2002)
  [arXiv:hep-th/0206255].
  %%CITATION = HEP-TH 0206255;%%
}

%\BerkovitsVY
\lref\BerkovitsVY{
  N.~Berkovits and C.~Vafa,
  ``N=4 topological strings,''
  Nucl.\ Phys.\ B {\bf 433}, 123 (1995)
  [arXiv:hep-th/9407190].
  %%CITATION = HEP-TH 9407190;%%
}

%\CheungYW
\lref\CheungYW{
  Y.~K.~Cheung, Y.~Oz and Z.~Yin,
  ``Families of N = 2 strings,''
  JHEP {\bf 0311}, 026 (2003)
  [arXiv:hep-th/0211147].
  %%CITATION = HEP-TH 0211147;%%
}

%\WittenMK
\lref\WittenMK{
  E.~Witten,
  ``The N matrix model and gauged WZW models,''
  Nucl.\ Phys.\ B {\bf 371}, 191 (1992).
  %%CITATION = NUPHA,B371,191;%%
}

%\LiKE
\lref\LiKE{
  K.~Li,
  ``Topological Gravity With Minimal Matter,''
  Nucl.\ Phys.\ B {\bf 354}, 711 (1991).
  %%CITATION = NUPHA,B354,711;%%
}

%\NakamuraSM
\lref\NakamuraSM{
S.~Nakamura and V.~Niarchos,
``Notes on the S-matrix of bosonic and topological non-critical strings,''
arXiv:hep-th/0507252.
%%CITATION = HEP-TH 0507252;%%
}

%\KutasovXU
\lref\KutasovXU{
  D.~Kutasov and N.~Seiberg,
  ``More comments on string theory on AdS(3),''
  JHEP {\bf 9904}, 008 (1999)
  [arXiv:hep-th/9903219].
  %%CITATION = HEP-TH 9903219;%%
}

%\NeitzkeNI
\lref\NeitzkeNI{
  A.~Neitzke and C.~Vafa,
  ``Topological strings and their physical applications,''
  arXiv:hep-th/0410178.
  %%CITATION = HEP-TH 0410178;%%
}

%\SahakyanDH
\lref\SahakyanDH{
  D.~A.~Sahakyan and T.~Takayanagi,
  ``On the Connection between N=2 Minimal String and (1,n) Bosonic Minimal
  String,''
  arXiv:hep-th/0512112.
  %%CITATION = HEP-TH 0512112;%%
}

%\RastelliPH
\lref\RastelliPH{
  L.~Rastelli and M.~Wijnholt,
  ``Minimal AdS(3),''
  arXiv:hep-th/0507037.
  %%CITATION = HEP-TH 0507037;%%
}

%\TakayanagiYJ
\lref\TakayanagiYJ{
  T.~Takayanagi,
  ``Notes on S-matrix of non-critical N = 2 string,''
  JHEP {\bf 0509}, 001 (2005)
  [arXiv:hep-th/0507065].
  %%CITATION = HEP-TH 0507065;%%
}

%\DijkgraafDH
\lref\DijkgraafDH{
  R.~Dijkgraaf and C.~Vafa,
  ``A perturbative window into non-perturbative physics,''
  arXiv:hep-th/0208048.
  %%CITATION = HEP-TH 0208048;%%
}

%\BershadskyCX
\lref\BershadskyCX{
  M.~Bershadsky, S.~Cecotti, H.~Ooguri and C.~Vafa,
  ``Kodaira-Spencer theory of gravity and exact results for quantum string
  amplitudes,''
  Commun.\ Math.\ Phys.\  {\bf 165}, 311 (1994)
  [arXiv:hep-th/9309140].
  %%CITATION = HEP-TH 9309140;%%
}

%\AntoniadisZE
\lref\AntoniadisZE{
  I.~Antoniadis, E.~Gava, K.~S.~Narain and T.~R.~Taylor,
  ``Topological amplitudes in string theory,''
  Nucl.\ Phys.\ B {\bf 413}, 162 (1994)
  [arXiv:hep-th/9307158].
  %%CITATION = HEP-TH 9307158;%%
}

%\OoguriZV
\lref\OoguriZV{
  H.~Ooguri, A.~Strominger and C.~Vafa,
  ``Black hole attractors and the topological string,''
  Phys.\ Rev.\ D {\bf 70}, 106007 (2004)
  [arXiv:hep-th/0405146].
  %%CITATION = HEP-TH 0405146;%%
}

%\OoguriWW
\lref\OoguriWW{
  H.~Ooguri and C.~Vafa,
  ``Selfduality And N=2 String Magic,''
  Mod.\ Phys.\ Lett.\ A {\bf 5}, 1389 (1990).
  %%CITATION = MPLAE,A5,1389;%%
}

%\GiveonPX
\lref\GiveonPX{
  A.~Giveon and D.~Kutasov,
  ``Little string theory in a double scaling limit,''
  JHEP {\bf 9910}, 034 (1999)
  [arXiv:hep-th/9909110].
  %%CITATION = HEP-TH 9909110;%%
}

%\KiritsisSS
\lref\KiritsisSS{
  E.~Kiritsis,
  ``Duality and instantons in string theory,''
  arXiv:hep-th/9906018.
  %%CITATION = HEP-TH 9906018;%%
}

%\BachasMC
\lref\BachasMC{
  C.~Bachas, C.~Fabre, E.~Kiritsis, N.~A.~Obers and P.~Vanhove,
  ``Heterotic/type-I duality and D-brane instantons,''
  Nucl.\ Phys.\ B {\bf 509}, 33 (1998)
  [arXiv:hep-th/9707126].
  %%CITATION = HEP-TH 9707126;%%
}

%\StiebergerWK
\lref\StiebergerWK{
  S.~Stieberger and T.~R.~Taylor,
  ``Non-Abelian Born-Infeld action and type I - heterotic duality. II:
  Nonrenormalization theorems,''
  Nucl.\ Phys.\ B {\bf 648}, 3 (2003)
  [arXiv:hep-th/0209064].
  %%CITATION = HEP-TH 0209064;%%
}

%\KostovIE
\lref\KostovIE{
  I.~K.~Kostov,
  ``Gauge invariant matrix model for the A-D-E closed strings,''
  Phys.\ Lett.\ B {\bf 297}, 74 (1992)
  [arXiv:hep-th/9208053].
  %%CITATION = HEP-TH 9208053;%%
}

%\EguchiVZ
\lref\EguchiVZ{
  T.~Eguchi and S.~K.~Yang,
  ``N=2 Superconformal Models As Topological Field Theories,''
  Mod.\ Phys.\ Lett.\ A {\bf 5}, 1693 (1990).
  %%CITATION = MPLAE,A5,1693;%%
}

%\DijkgraafNC
\lref\DijkgraafNC{
  R.~Dijkgraaf and E.~Witten,
  ``Mean Field Theory, Topological Field Theory, And Multimatrix Models,''
  Nucl.\ Phys.\ B {\bf 342}, 486 (1990).
  %%CITATION = NUPHA,B342,486;%%
}

%\DijkgraafQH
\lref\DijkgraafQH{
  R.~Dijkgraaf,
  ``Intersection theory, integrable hierarchies and topological field theory,''
  arXiv:hep-th/9201003.
  %%CITATION = HEP-TH 9201003;%%
}

%\KontsevichTI
\lref\KontsevichTI{
  M.~Kontsevich,
  ``Intersection theory on the moduli space of curves and the matrix Airy
  function,''
  Commun.\ Math.\ Phys.\  {\bf 147}, 1 (1992).
  %%CITATION = CMPHA,147,1;%%
}

%\KiritsisPB
\lref\KiritsisPB{
  E.~Kiritsis, C.~Kounnas and D.~Lust,
  ``A Large class of new gravitational and axionic backgrounds for
  four-dimensional superstrings,''
  Int.\ J.\ Mod.\ Phys.\ A {\bf 9}, 1361 (1994)
  [arXiv:hep-th/9308124].
  %%CITATION = HEP-TH 9308124;%%
}

%\AdemolloPP
\lref\AdemolloPP{
  M.~Ademollo {\it et al.},
  ``Dual String With U(1) Color Symmetry,''
  Nucl.\ Phys.\ B {\bf 111}, 77 (1976).
  %%CITATION = NUPHA,B111,77;%%
}

%\MarcusWI
\lref\MarcusWI{
  N.~Marcus,
  ``A Tour through N=2 strings,''
  arXiv:hep-th/9211059.
  %%CITATION = HEP-TH 9211059;%%
}

%\GiribetIX
\lref\GiribetIX{
  G.~Giribet and Y.~Nakayama,
  ``The Stoyanovsky-Ribault-Teschner map and string scattering amplitudes,''
  arXiv:hep-th/0505203.
  %%CITATION = HEP-TH 0505203;%%
}

%\BershadskyPE
\lref\BershadskyPE{
  M.~Bershadsky, W.~Lerche, D.~Nemeschansky and N.~P.~Warner,
  ``Extended N=2 superconformal structure of gravity and W gravity coupled to
  matter,''
  Nucl.\ Phys.\ B {\bf 401}, 304 (1993)
  [arXiv:hep-th/9211040].
  %%CITATION = HEP-TH 9211040;%%
}

%\AshokXC
\lref\AshokXC{
  S.~K.~Ashok, S.~Murthy and J.~Troost,
  ``Topological cigar and the c = 1 string: Open and closed,''
  arXiv:hep-th/0511239.
  %%CITATION = HEP-TH 0511239;%%
}

%\TakayanagiYB
\lref\TakayanagiYB{
  T.~Takayanagi,
  ``c$<$1 string from two dimensional black holes,''
  JHEP {\bf 0507}, 050 (2005)
  [arXiv:hep-th/0503237].
  %%CITATION = HEP-TH 0503237;%%
}

%\ZamolodchikovBD
\lref\ZamolodchikovBD{
  A.~B.~Zamolodchikov and V.~A.~Fateev,
  ``Operator Algebra And Correlation Functions In The Two-Dimensional
  Wess-Zumino SU(2) X SU(2) Chiral Model,''
  Sov.\ J.\ Nucl.\ Phys.\  {\bf 43}, 657 (1986)
  [Yad.\ Fiz.\  {\bf 43}, 1031 (1986)].
  %%CITATION = SJNCA,43,657;%%
}

%\TeschnerFT
\lref\TeschnerFT{
  J.~Teschner,
  ``On structure constants and fusion rules in the SL(2,C)/SU(2) WZNW  model,''
  Nucl.\ Phys.\ B {\bf 546}, 390 (1999)
  [arXiv:hep-th/9712256].
  %%CITATION = HEP-TH 9712256;%%
}

%\TeschnerUG
\lref\TeschnerUG{
  J.~Teschner,
  ``Operator product expansion and factorization in the H-3+ WZNW model,''
  Nucl.\ Phys.\ B {\bf 571}, 555 (2000)
  [arXiv:hep-th/9906215].
  %%CITATION = HEP-TH 9906215;%%
}

%\ItzhakiTU
\lref\ItzhakiTU{
  N.~Itzhaki, D.~Kutasov and N.~Seiberg,
  ``I-brane dynamics,''
  arXiv:hep-th/0508025.
  %%CITATION = HEP-TH 0508025;%%
}

%\IsraelIR
\lref\IsraelIR{
  D.~Israel, C.~Kounnas, A.~Pakman and J.~Troost,
  ``The partition function of the supersymmetric two-dimensional black hole
  and little string theory,''
  JHEP {\bf 0406}, 033 (2004)
  [arXiv:hep-th/0403237].
  %%CITATION = HEP-TH 0403237;%%
}

%\Rastelli
\lref\Rastelli{
L.~Rastelli,
``Minimal superstrings, topological strings and NSF5 branes,''
talk given at the Simons workshop in Mathematics and Physics 2005. 
The talk can be found online at http://insti.physics.sunysb.edu/itp/conf/simonswork3/
}

%%%%%%%%%%%%%%%%%%%%%%
%                      Title Page                                 %
%%%%%%%%%%%%%%%%%%%%%%
\Title{
%\vbox{\hbox{EFI-03-41}
%      \hbox{hep-th/0308172}}
}
{\vbox{
\centerline{On Minimal $\NN=4$ Topological Strings}
\vskip0.4cm
\centerline{And The $(1,k)$ Minimal Bosonic String}}}

\vskip .2in

\centerline{Vasilis Niarchos\footnote{$^{\dagger}$}{niarchos@nbi.dk}}

\vskip .2in

%\vskip 2cm
\centerline{{\it The Niels Bohr Institute}}
\centerline{\it Blegdamsvej 17, 2100 Copenhagen \O, Denmark}

\vskip 2cm
\noindent

%%abstract

In this paper we consider tree-level scattering in 
the minimal $\NN=4$ topological string and show that 
a large class of $N$-point functions can be recast in terms 
of corresponding amplitudes in the $(1,k)$ minimal bosonic string. 
This suggests a non-trivial
relation between the minimal $\NN=4$ topological strings, the $(1,k)$
minimal bosonic strings and their corresponding ADE matrix models.
This relation has interesting and far-reaching 
implications for the topological sector of six-dimensional Little String
Theories.

\Date{December, 2005}

\vfill
\vfill

\listtoc
\writetoc

\newsec{Introduction}

The properties of topological strings have been studied over the years
in much detail and by now it is fairly clear that topological strings are much
more than a toy model for string theory (for a recent review see \NeitzkeNI). 
Indeed, many computations in topological string theory 
have been shown to have a direct consequence for ten-dimensional 
superstrings. For example, the $\NN=2$ topological string has played a role
in the computation of low-energy superpotential terms in 
four-dimensional $\NN=1$ supersymmetric gauge
theories and has led to an interesting connection between gauge theory 
effective superpotentials and zero-dimensional bosonic matrix models 
\refs{\DijkgraafFC,\DijkgraafDH}. Another application involves
type II superstring theory compactified on a Calabi-Yau threefold. The four-dimensional
low-energy effective action includes higher-derivative $F$-type terms of the 
form
\eqn\aaa{
\int d^4 x d^4\theta(W_{ab}W^{ab})^g F_g(X^{\Lambda})
~,}
where $W_{ab}$ is the graviphoton superfield of the $\NN=2$ supergravity theory
and $X^{\Lambda}$ are vector multiplets. The $\NN=2$ topological string computes
the functions $F_g$ \refs{\AntoniadisZE,\BershadskyCX}. In recent developments, 
this fact has been used to formulate an intriguing relation between the topological 
string partition function and the partition function of dyonic BPS black holes 
in four-dimensional $\NN=2$ supergravity \OoguriZV.

In the past, much work has been done on the $\NN=2$ topological string theory.
In this case a worldsheet theory with $\NN=(2,2)$ supersymmetry 
is topologically twisted and the critical central charge is $\hat c=\frac{c}{3}=3$.
There are, however, other interesting examples where the amount of worldsheet supersymmetry
and central charge are different and where a non-tirivial 
topological string theory can be defined. Most notably, worldsheet theories with
$\NN=(4,4)$ supersymmetry and $\hat c=2$ can be twisted appropriately
to obtain the $\NN=4$ topological string. This theory has been argued \BerkovitsVY\ to be 
equivalent to the $\NN=2$ string,\foot{Not to be confused with the $\NN=2$ topological
string.} which is obtained when the full $\NN=2$
superconformal group is gauged on the worldsheet 
\refs{\AdemolloPP,\OoguriWW,\OoguriFP,\MarcusWI}. 
In the same way that the $\NN=2$ topological string is relevant for type II
string theory compactifications on Calabi-Yau threefolds, the $\NN=4$ topological
string (or $\NN=2$ string) is directly relevant for type II string theory compactifications
on K3 manifolds. The coefficient of certain terms of the form $R^4F^{4g-4}$ 
in the low-energy effective action of the latter compactifications can be computed in the $\NN=4$ topological string \BerkovitsVY. The coefficient of a term of the form $F^4$, where
$F$ is an abelian gauge field in the low-energy theory, was the focus of the analysis
of \AharonyVK.

It is known \refs{\OoguriWJ,\GiveonZM,\GiveonPX} that the dynamics of string 
theory near an ADE singularity of a K3 compactification are captured 
by an asymptotically linear dilaton theory of the form 
\eqn\aab{
\IR^{5,1}\times \bigg(\frac{SL(2)_k}{U(1)}\times \frac{SU(2)_k}{U(1)}\bigg)/\IZ_k
~}
and that this background defines holographically a Little String Theory (LST)
(for a review see \refs{\AharonyKS,\KutasovUF}).
Via T-duality this background is also related to the near-horizon region
of $k$ parallel NS5-branes symmetrically arranged on a transverse circle.
We can study the topological sector of these six-dimensional LST's by
analyzing the $\NN=4$ topological string on 
$\big(\frac{SL(2)_k}{U(1)}\times \frac{SU(2)_k}{U(1)}\big)/\IZ_k$.
For obvious reasons, we will call this theory the minimal $\NN=4$ topological
string. Earlier work on this theory appeared in \refs{\AharonyVK,\KonechnyDF}.

In this paper, we compute a large class of tree-level correlation functions
in the above theory. In contrast to the $\NN=4$ topological string on $\IR^4$ (or
$\IR^{2,2}$), which has vanishing $N$-point functions for $N\geq 4$ \OoguriWW, 
we will see that the minimal
$\NN=4$ topological string has non-vanishing amplitudes for any $N$.
In fact, using recent results on tree-level correlation functions in
$SL(2,\IR)$ \refs{\StoyanovskyPG,\RibaultWP, \GiribetIX,\RibaultMS} 
we can recast in many cases the topological string $N$-point function
in terms of a corresponding $N$-point function in the $(1,k)$ minimal bosonic
string. Among other things, this allows us to reformulate as a four-point
function in the bosonic string, a four-point function that appeared in \AharonyVK\ 
and gives the coefficient of an $F^4$ term
in the low-energy effective action of a K3 compactification near an ADE point
of enhanced symmetry. We expect this observation 
to be particularly useful in completing the non-trivial check of heterotic-type II duality 
presented in \AharonyVK\ (related discussions on this $F^4$ term can 
be found in \refs{\BachasMC,\KiritsisSS,\StiebergerWK}).
 
The emergence of a bosonic string in the context of the minimal $\NN=4$
topological string reminds of the well-known
equivalence between the $\NN=2$ topological string on $SL(2)_1/U(1)$
and the bosonic $c=1$ string at self-dual radius \refs{\MukhiZB,\OhtaEH,\GhoshalWM}.\foot{See
also the recent discussions in \refs{\TakayanagiYB,\NakamuraSM,\AshokXC}.}
In this paper we find evidence pointing towards a non-trivial relation between
the minimal bosonic string and the minimal $\NN=4$ topological string.\foot{A relation
between the $(1,k)$ minimal bosonic string and the $A_k$ minimal $\NN=4$ topological
string has been proposed also by Leonardo Rastelli in \Rastelli\ based
on work in collaboration with Martijn Wijnholt.}
It is natural to ask whether this relation is a true equivalence. 
This would have a non-trivial impact on the physics of six-dimensional LST's. 
For example, it would imply that the topological sector of these (yet quite mysterious) theories 
is completely solved by the ADE matrix models of ref.\ \KostovIE. 
Some comments on this possible equivalence will be presented in the last section. 

The organization of this paper is as follows. In section $2$ we fix our notation
and discuss the spectrum of the minimal $\NN=4$ topological string.
In section $3$ we compute a large class of $N$-point functions
in the topological string and recast them in terms of corresponding amplitudes
in the $(1,k)$ minimal bosonic string with the use of the Stoyanovsky-Ribault-Teschner
(SRT) map. In section $4$ we comment on the possibility of an equivalence between
the minimal $\NN=4$ topological string and the $(1,k)$ minimal bosonic string
and point out some implications. Four appendices contain useful facts and 
summarize our conventions.

\vskip 0.5cm
\noindent
{\bf Note added}: While this paper was being prepared, we received
the preprint \SahakyanDH, which has considerable overlap with this work.
Our analysis deals with the same amplitudes and yields similar results, 
but is based on a slightly different treatment of the 
$\frac{SL(2)}{U(1)}\times \frac{SU(2)}{U(1)}$ CFT. 

\

\

\newsec{Minimal $\NN=4$ topological strings - the spectrum}

\subsec{Fixing the notation}

The minimal $\NN=4$ topological string is defined
by twisting topologically (in a manner that will be reviewed
shortly) the $\NN=4$ superconformal field theory \KiritsisPB
\eqn\baa{
\bigg(\frac{SL(2)_k}{U(1)}\times \frac{SU(2)_k}{U(1)}\bigg)/\IZ_k
~.}
The total central charge of this theory is 
\eqn\bab{
\hat c_{\rm tot}=1+\frac{2}{k}+1-\frac{2}{k}=2
~.}
As required, this is the critical value for the $\NN=4$
topological string. The $\IZ_k$ orbifold in \baa\ restricts
the spectrum to the states with integral $U(1)_R$ charge
as needed by modular invariance. The conformal field theory \baa\
can be obtained from $SL(2)_k\times SU(2)_k$ with a
suitable $U(1)$ gauging. We summarize briefly a few of
the relevant details to fix the notation.

The $\NN=1$ superconformal field theory on $SU(2)_k$ is
given by the bosonic $SU(2)_{k-2}$ WZW model with currents
$K^a$, $a=1,2,3$ and three free Majorana fermions $\chi^a$. The
relevant OPE's are
\eqn\bac{\eqalign{
K^3(z)K^3(w)\sim \frac{k-2}{2(z-w)^2}~, ~
K^+(z)&K^-(w)\sim \frac{k-2}{(z-w)^2}+\frac{2K^3(w)}{z-w}~, 
\cr
K^3(z)K^{\pm}(w)\sim& \pm \frac{K^{\pm}(w)}{z-w}
~,}}
\eqn\bad{
\chi^+(z)\chi^-(w)\sim \frac{1}{z-w}~, ~ ~
\chi^3(z)\chi^3(w)\sim \frac{1}{z-w}
~.}
In terms of these currents the $\NN=1$ superconformal generator 
takes the form
\eqn\bae{
G^{(su)}=Q\Big(\frac{1}{\sqrt 2}K^+\chi^-+\frac{1}{\sqrt 2} K^-\chi^+
+K^3\chi^3+\chi^+\chi^-\chi^3\Big)
~,}
where $Q=\sqrt{\frac{2}{k}}$.

We will use a similar set of conventions for the supersymmetric
$SL(2)_k$ WZW model. This theory consists of the bosonic
$SL(2)_{k+2}$ WZW model with currents $J^a$ ($a=1,2,3$)
and three free fermions $\psi^a$, which satisfy the OPE algebra
\eqn\baf{\eqalign{
J^3(z)J^3(w)\sim -\frac{k+2}{2(z-w)^2}~, ~
J^+(z)&J^-(w)\sim \frac{k+2}{(z-w)^2}-\frac{2J^3(w)}{z-w}~, 
\cr
J^3(z)J^{\pm}(w)&\sim \pm \frac{J^{\pm}(w)}{z-w}
~,}}
\eqn\bag{
\psi^+(z)\psi^-(w)\sim \frac{1}{z-w}~, ~ ~
\psi^3(z)\psi^3(w)\sim -\frac{1}{z-w}
~.}
The $SL(2)$ $\NN=1$ supercurrent is 
\eqn\bai{
G^{(sl)}=Q\Big(\frac{1}{\sqrt 2}J^+\psi^-+\frac{1}{\sqrt 2} J^-\psi^+
-J^3\psi^3-\psi^+\psi^-\psi^3\Big)
~.}

The product CFT \baa\ can be obtained from the supersymmetric 
$SL(2)_k\times SU(2)_k$ theory by gauging the null $U(1)$ supercurrent
(see \refs{\IsraelIR,\ItzhakiTU})
\eqn\baj{
\frac{1}{\sqrt 2}(\chi^3-\psi^3)-\theta \frac{Q}{\sqrt 2}(J_3^{(tot)}-K_3^{(tot)})
~,}
where by definition
\eqn\bak{
J_3^{(tot)}=J_3+\psi^+\psi^-~, ~ ~ K_3^{(tot)}=K_3+\chi^+\chi^-
~.}
As usual, the gauging of the supercurrent \baj\ can be achieved by adding
a bosonic $(\beta',\gamma')$ ghost system with spin $\frac{1}{2}$,
which contributes to the total BRST charge a term of the form
\eqn\bal{
Q_{BRST}=...+\int \frac{dz}{2\pi i} \gamma'(z)\frac{1}{\sqrt 2}(\chi^3-\psi^3)
~.}
Then one can check up to BRST exact terms that the $\NN=1$  supercurrent
of the gauged theory is
\eqn\bam{
G=G^{(su)}+G^{(sl)}=\frac{Q}{\sqrt 2}(J^+\psi^-+J^-\psi^++K^+\chi^-+K^-\chi^+)
~.}

It turns out that the gauged theory enjoys enhanced $\NN=(4,4)$
worldsheet supersymmetry. The left-moving $\NN=2$ 
superconformal generators take the form
\eqn\ban{\eqalign{
G^{\pm}&=\frac{Q}{\sqrt 2}\Big(e^{\mp i H_1}J^{\pm}+e^{\mp i H_2}K^{\pm}\Big)~,
\cr
J&=-i\d H_1-i\d H_2
~,}}
where $H_1$ and $H_2$ are two bosons that bosonize the free fermions
$\psi^{\pm}$, $\chi^{\pm}$, $i.e.$
\eqn\bao{
\psi^{\pm}=e^{\pm i H_1}~, ~ ~ \chi^{\pm}=e^{\pm i H_2}
~.}
For the right-moving sector we use the convention
\eqn\bap{\eqalign{
\bar G^{\pm}&=\frac{Q}{\sqrt 2}\Big(e^{\pm i \bar H_1}\bar J^{\mp}
+e^{\pm i \bar H_2}\bar K^{\mp}\Big)~,
\cr
\bar J&=i\bar \d H_1+i\bar \d H_2
~.}}
The theory includes additional superconformal generators. They read
\eqn\bba{
J^{\pm \pm}=e^{\pm \int J}=e^{\mp i (H_1+H_2)}~, ~ ~
\bar J^{\pm \pm}=e^{\pm \int \bar J}=e^{\pm i (\bar H_1+\bar H_2)}
~,}
\eqn\bbb{
\tilde G^{\pm}=\frac{Q}{\sqrt 2}\big(e^{\mp i H_2}J^{\mp}+
e^{\mp i H_1}K^{\mp}\big)~, ~ ~
\bar {\tilde G}^{\pm}=\frac{Q}{\sqrt 2}\big(e^{\pm i \bar H_2}\bar J^{\pm}+
e^{\pm i \bar H_1}\bar K^{\pm}\big)
~.}
One can check that the full set of currents $T(z), J(z), J^{\pm \pm}(z), G^{\pm}(z), 
\tilde G^{\pm}(z)$ satisfies the $\NN=4$ superconformal algebra 
(see appendix A for more details on this algebra). The $\pm$ superscripts in 
\bba, \bbb\ denote the $U(1)_R$ charge of the corresponding generators.
For example, the currents $J^{\pm \pm}$ have $U(1)_R$ charges $\pm 2$ respectively and 
$\tilde G^{\pm}$ have $U(1)_R$ charges $\pm 1$.

\subsec{Physical states in $SL(2)/U(1)\times SU(2)/U(1)$}

Physical states in $\frac{SL(2)}{U(1)}\times \frac{SU(2)}{U(1)}$
can be obtained by imposing the $U(1)$ gauging condition to a
general vertex operator in $SL(2)\times SU(2)$. For example,
consider the special set of vertex operators
\eqn\bca{
\VV=e^{i\sum_{i=1}^2 (s_i H_i+\bar s_i \bar H_i)}
\Phi^{(sl)}_{j_1,m_1,\bar m_1} \Phi^{(su)}_{j_2,m_2,\bar m_2}
~,} 
where $\Phi^{(sl)}_{j,m,\bar m}$, $\Phi^{(su)}_{j,m,\bar m}$ are  
respectively primary vertex operators of the bosonic $SL(2)_{k+2}$ 
and $SU(2)_{k-2}$ WZW models. The relevant conventions are summarized in 
appendix B. The null gauging condition $K_3^{(tot)}=J_3^{(tot)}$
implies the constraints
\eqn\bcb{
m_2+s_2=m_1+s_1~, ~ ~ \bar m_2+\bar s_2=\bar m_1+\bar s_1
~.}
For reference, we mention that the primary fields \bca\ have 
scaling dimension
\eqn\bcc{
\Delta=\frac{1}{2}(s_1^2+s_2^2)-\frac{j_1(j_1+1)}{k}+
\frac{j_2(j_2+1)}{k}
}
and $U(1)_R$ charge
\eqn\bcd{
q=-s_1-s_2
~.}

In the context of the topological string a special role will be
played by the NS sector chiral primary states. In particular,
these states have the property $\Delta=\frac{q}{2}$ \LercheUY. For the 
operators \bca\ this is the case when
\eqn\bce{
\frac{1}{2}(s_1^2+s_2^2)-\frac{j_1(j_1+1)}{k}+
\frac{j_2(j_2+1)}{k}=-\frac{1}{2}(s_1+s_2)
~.}
For general level $k$ and quantum numbers $j$
\bce\ is satisfied if $j_1=j_2$ or $-j_1-1=j_2$ and
\eqn\bcf{
s_1^2+s_2^2=-s_1-s_2
~.}
The last equation holds in the following four cases
\eqn\bcg{
(s_1,s_2)=\{ (0,0),(0,-1),(-1,0),(-1,-1) \}
~.}
Analogous statements apply to the right-moving sector
and with obvious modifications to the anti-chiral states.

\subsec{Physical states in the topological theory}

A theory with $\NN=4$ superconformal symmetry can
be twisted topologically in the same way as a theory
with $\NN=2$ superconformal symmetry. Once we choose a 
$U(1)$ inside the $SU(2)$ R-symmetry group we can perform
the familiar $\NN=2$ topological twist
\eqn\bda{
T\rightarrow T+\frac{1}{2}\d J~,
~ ~\bar T\rightarrow \bar T\pm \frac{1}{2}\bar \d \bar J
~.}
As in the $\NN=2$ topological string, it is possible 
to make a choice of two inequivalent topological twists: the
A-type corresponds to the minus sign in \bda\ and the B-type  
to the plus sign. Each of these choices is expected to
be in one-to-one correspondence with an $\alpha$- or $\beta$-type $\NN=2$
string \CheungYW. For concreteness, in this paper we will 
consider the A-type topological twist.

The next and more crucial step is to decide how to define the
BRST cohomology of the topologically twisted theory and how
to couple the twisted theory to gravity. For the $\NN=2$ topological
string it is enough to consider the cohomology of $(c,c)$ or
$(c,a)$ fields $\phi_i$. For the $\NN=4$ topological string the BRST cohomology
is more constrained \BerkovitsVY\ and can be obtained by further restricting
the fields $\phi_i$ to have the properties\foot{These relations define
the appropriate BRST cohomology for the A-type topological twist. Similar
relations define the BRST cohomology of the B-type topological string.}
\eqn\bdb{
G^+\phi_i=0~, ~ ~ \tilde G^+\phi_i=0~, ~ ~
\bar G^- \phi_i=0~, ~ ~ \bar{\tilde G}^-\phi_i=0
~}
and the equivalence relation
\eqn\bdc{
\phi_i\sim \phi_i+ G^+\tilde G^+\bar G^-\bar{\tilde G}^-\chi
~.}
In addition, one has to impose the properties
\eqn\bdd{\eqalign{
J^{--}\phi_i&=M_i^{\bar j} \phi^{\dagger}_{\bar j}~,
~ ~ J^{++}\phi_i=0~,
\cr
\bar J^{++}\phi_i&=\bar M^{\bar j}_i \bar \phi^{\dagger}_{\bar j}~,
~ ~ \bar J^{--}\bar \phi_i=0
~.}}
The first and third equations in \bdd\ impose the requirement that 
the fields $J^{--}\phi_i$ ($\bar J^{++}\phi_i$) can be expressed
as a linear combination of antichiral (chiral) fields. $M_i^{\bar j}$ and
$\bar M_i^{\tilde j}$ are the linear coefficients in this expansion.

For the vertex operators $\VV$ in \bca\ the above 
requirements can be imposed in the following way
(here we discuss the left-movers, but analogous statements
apply also to the right-movers). The first condition 
in \bdb\ is automatic for the chiral primary states.
The second condition gives
\eqn\bde{\eqalign{
&\int dz~\tilde G^+(z)\cdot \VV(0) =0 ~ \Leftrightarrow ~
\int dz ~\big(e^{-iH_2}J^-+e^{-iH_1}K^-\big)(z)\cdot \VV(0) =0 \Leftrightarrow
\cr
&\int dz~ \Big(z^{-s_2-1} (-j_1-1+m_1) e^{-iH_2(z)+is_2H_2(0)}
\Phi^{(sl)}_{j_1,m_1-1,\bar m_1}(0)\cdots \Big)+
\cr
&\int dz~ \Big(z^{-s_1-1}(j_2+m_2)e^{-iH_1(z)+is_1H_1(0)}
\Phi^{(su)}_{j_2,m_2-1,\bar m_2}(0)
\cdots \Big)=0
~.}}
In this expression, the dots refer to the parts of the vertex operator
$\VV$, which are unaffected by the action of $\tilde G^+$.
To derive \bde\ we made use of the OPE's
\eqn\bdf{
J^-(z)~\Phi^{(sl)}_{j,m}(w)\sim \frac{-j-1+m}{z-w}~\Phi^{(sl)}_{j,m-1}(w)
~,}
\eqn\bdg{
K^-(z)~\Phi^{(su)}_{j,m}(w)\sim \frac{j+m}{z-w}~\Phi^{(su)}_{j,m-1}(w)
~.}
The perhaps unorthodox factor $-j-1+m$ (instead of $j+m$) in
\bdf\ has its origin in our $SL(2)$ conventions, which are reviewed
in appendix B. In a similar fashion, in order to check the first condition
in \bdd\ we need to compute
\eqn\bdi{
\int dz~ J^{--}(z)\cdot \VV(0)=
\int dz~ z^{s_1+s_2} e^{i(H_1+H_2)(z)+i(s_1H_1+is_2H_2)(0)}\cdots
~,}
which should be expressible as a sum of antichiral primary fields.
For the second condition in \bdd\ we get
\eqn\bdj{
\int dz~ J^{++}(z)\cdot \VV(0) = \int dz~ z^{-s_1-s_2} 
e^{-i(H_1+H_2)(z)+i(s_1H_1+s_2H_2)(0)}\cdots=0
~.}

Out of the four different cases in \bcg\ the $(s_1,s_2)=(-1,-1)$ case
fails to satisfy the condition $J^{--}\cdot\VV_i=M^{\bar j}_i\VV^{\dagger}_j$.
Indeed, \bdi\ gives
\eqn\bdk{
J^{--}\cdot e^{-H_1-H_2}\Phi^{(sl)}_{j,m}\Phi^{(su)}_{j,m}
=i(\d H_1+i\d H_2)\Phi^{(sl)}_{j,m}\Phi^{(su)}_{j,m}
~,}
which is not an antichiral field. The remaining three cases
$(s_1,s_2)=\{ (0,0),(-1,0),(0,-1)\}$ satisfy the full set of conditions
provided that we tune the quantum numbers $j_i,m_i$ appropriately.
We conclude that the BRST cohomology of the minimal $\NN=4$ topological 
string contains (at least) the following states
\eqn\bdl{\eqalign{
&\OO_{2j+1}=\Phi^{(sl)}_{-j-1,-j,-j}\Phi^{(su)}_{j,-j,-j}~, ~ ~
\cr
\OO^+_{2j+1}=e^{-iH_1-i\bar H_1}\Phi^{(sl)}_{j,j+1,j+1}&\Phi^{(su)}_{j,j,j}~,~ ~
\OO^-_{2j+1}=e^{-iH_2-i \bar H_2}\Phi^{(sl)}_{j,-j-1,-j-1}\Phi^{(su)}_{j,-j,-j}
~.}}
$j$ is a half-integer with the property $0\leq j\leq \frac{k-2}{2}$.

The vertex operators $\OO^{\pm}_{2j+1}$ are normalizable in 
$\frac{SL(2)}{U(1)}\times \frac{SU(2)}{U(1)}$ (see appendix B for more 
details) and according to the
general discussion in \AharonyVK\ their correlation functions compute
amputated amplitudes in the holographic non-gravitational dual (see
also \AharonyXN).
These correlation functions will be computed in the next section.
In addition, it will be useful to note that the vertex operators
$\OO^{+}_{2j+1}$ and $\OO^-_{2j+1}$ are related via the reflection
relation
\eqn\bdm{
\OO^+_{2j+1}=\OO^-_{k-2j-1}
~.}
This is shown explicitly in appendix B.

In contrast, the zero $R$-charge vertex operators $\OO_{2j+1}$
are non-normalizable. It is worth pointing out, however, that 
the normalizable version of $\OO_{2j+1}$ is related to the
$\OO^-_{2j+1}$ state via the relation
\eqn\bdn{
\OO^-_{2j+1}=\frac{k}{(2j+1)^2}\tilde G^+ \bar{\tilde G}^-\cdot 
\Big(\Phi^{(sl)}_{j,-j,-j}\Phi^{(su)}_{j,-j,-j}\Big)
~.}

\subsec{Vertex operators in the $\NN=2$ string}

Here we make a short parenthesis to discuss the corresponding
vertex operators in the minimal $\NN=2$ string and to make contact with the
analysis of \AharonyVK. Concentrating only on the left-moving sector,
the physical $\NN=2$ string states $\hat \OO^{\pm}_{2j+1}$ that appear in \AharonyVK\
take in our notation the following form\foot{We drop
an appropriate normalization factor. Also $H$, $H'$ in \AharonyVK\ correspond
to $-H_1$, $-H_2$ in our notation and $j_{there}= -j_{here}-1$.}
\eqn\bea{\eqalign{
&\hat \OO^+_{2j+1}=e^{-\frac{1}{2}(\varphi_1+\varphi_2)+\frac{i}{2}(-H_1+H_2)}
\Phi^{(sl)}_{j,j+1}\Phi_{j,j}^{(su)}~,
\cr
&\hat \OO^-_{2j+1}=e^{-\frac{1}{2}(\varphi_1+\varphi_2)+\frac{i}{2}(H_1-H_2)}
\Phi^{(sl)}_{j,-j-1}\Phi_{j,-j}^{(su)}
~.}}
$\varphi_1$ and $\varphi_2$ are the two superconformal ghosts of the
$\NN=2$ string associated with the generators $G^-$ and $G^+$ respectively.
The operators appearing in \bea\ are in the R sector. They can be
converted to NS sector operators by using the spectral flow operation,
a transformation that is gauged in the $\NN=2$ string. The operators
that implement the spectral flow can be written as\foot{As in \AharonyVK\ we
omit a factor $e^{\pm \frac{1}{2}c\tilde b}$.}
\eqn\beaa{
S^{\pm}=e^{\pm\frac{1}{2}(\varphi_2-\varphi_1)\mp \frac{i}{2}(H_1+H_2)}
~.}
The basic properties of these operators are summarized in \AharonyVK\ 
(see pg.\ 44). Acting with them on the vertex operators $\hat \OO^+_{2j+1}$
we find
\eqn\beb{\eqalign{
S^-\hat\OO^+_{2j+1}&=e^{-\varphi_2+iH_2}\Phi^{(sl)}_{j,j+1}\Phi_{j,j}^{(su)}~, ~ ~
S^+\hat\OO^+_{2j+1}=e^{-\varphi_1-iH_1}\Phi^{(sl)}_{j,j+1}\Phi_{j,j}^{(su)}~,
\cr
S^-\hat\OO^-_{2j+1}&=e^{-\varphi_2+iH_1}\Phi^{(sl)}_{j,-j-1}\Phi_{j,-j}^{(su)}~, ~ ~
S^+\hat\OO^-_{2j+1}=e^{-\varphi_1-iH_2}\Phi^{(sl)}_{j,-j-1}\Phi_{j,-j}^{(su)}
~.}}
In the $\NN=2$ string there are two picture changing operators\foot{The 
ellipses denote a set of terms that depend only on ghosts.}
\eqn\bec{
Z^-=e^{\varphi_1}[G^-+...]~, ~ ~ 
Z^+=e^{\varphi_2}[G^++...]
~}
and we can use them to change the picture of the vertex operators \beb\ 
as follows 
\eqn\bed{\eqalign{
&S^+\hat \OO^+_{2j+1}\sim -\frac{\sqrt 2}{Q(2j+1)}e^{-\varphi_1-\varphi_2}
\Phi^{(sl)}_{j,j}\Phi^{(su)}_{j,j}~,
\cr
&S^-\hat \OO^-_{2j+1}\sim -\frac{\sqrt 2}{Q(2j+1)}e^{-\varphi_1-\varphi_2}
\Phi^{(sl)}_{j,-j}\Phi^{(su)}_{j,-j}
~.}}
We made use of the identities
\eqn\bee{\eqalign{
&G^+\cdot \Big(\Phi^{(sl)}_{j,j}\Phi^{(su)}_{j,j}\Big)=-\frac{Q}{\sqrt 2}(2j+1) 
e^{-iH_1}\Phi^{(sl)}_{j,j+1}\Phi^{(su)}_{j,j}~,
\cr
&G^-\cdot  \Big(\Phi^{(sl)}_{j,-j}\Phi^{(su)}_{j,-j}\Big)=-\frac{Q}{\sqrt 2}(2j+1) 
e^{iH_1}\Phi^{(sl)}_{j,-j-1}\Phi^{(su)}_{j,-j}
~.}}

As explained in \BerkovitsVY, vertex operators in the 
$\NN=2$ string of the form
\eqn\bef{
\hat \OO=c~ e^{-\varphi_1-\varphi_2}~V
}
correspond in the $\NN=4$ topological string to the $\tilde G^+$-exact vertex
operators
\eqn\beg{
\OO=\tilde G^+V
~.}
Combining eqs.\ \bdl, \bdn\ and \bed\ we deduce the 
correspondence
\eqn\bei{
\hat \OO^{-}_{2j+1} \leftrightarrow \OO^{-}_{2j+1}
~.}
The additional correspondence
\eqn\bej{
\hat \OO^{+}_{2j+1} \leftrightarrow \OO^{+}_{2j+1}
~}
follows from the $\NN=4$ topological string
reflection relation \bdm\ and the analogous relation in
the $\NN=2$ string $\hat \OO^+_{2j+1}=\hat \OO^-_{k-2j-1}$
(see eq.\ (5.34) in \AharonyVK).

\newsec{Minimal $\NN=4$ topological strings - $N$-point functions}

In this section we compute the tree-level correlation
functions of the vertex operators $\OO^{\pm}_{2j+1}$.
First, we consider the three- and four-point functions 
and then proceed with a general computation of $N$-point
functions for an arbitrary number of insertions $N\geq 5$.
Using the Stoyanovsky-Ribault-Teschner map
we show how to recast these correlation functions in terms of
corresponding amplitudes in the minimal $bosonic$ string.

\subsec{3-point functions}

In the $\NN=4$ topological string the three-point 
functions of $\tilde G^+\bar{\tilde G^-}$-exact operators 
$\phi_i=\tilde G^+\bar{\tilde G^-} V_i$ take the form \BerkovitsVY
\eqn\caa{
c_{ijk}=\langle \phi_i\phi_jV_k\rangle
~.}
With this definition the three-point function is symmetric in
the labels $(i,j,k)$ and respects all the invariances of the
$\NN=4$ topological string.
In this section, we will focus on correlation functions of
the vertex operators $\OO^-_{2j+1}$. A general amplitude
with both $\OO^+_{2j+1}$ and $\OO^-_{2j+1}$ insertions
follows trivially by using the reflection relation \bdm.

For reasons that will become more apparent in the following,
a vertex operator $\OO^-_{2j+1}$ will be inserted in the 
amplitude in the $U(1)$ gauge equivalent form\foot{A similar
trick was also used in the $\NN=2$ topological string
\refs{\TakayanagiYB, \NakamuraSM}.}
\eqn\cab{
\OO^-_{2j+1}\sim e^{\int (J_g+\bar J_g)}\OO^-_{2j+1}=
e^{-iH_1-i\bar H_1}\Phi^{(sl),w=1}_{j,-j-1,-j-1}
\Phi^{(su),w=1}_{j,-j,-j}
~.}
Recall that we gauge the null $U(1)$ current  
\eqn\caba{
J_g=i\d H_2-i\d H_1+K_3-J_3
~.}
This current commutes with all the generators of the $\NN=4$
superconformal algebra.
When it acts on $\OO^-_{2j+1}$ as in \cab, part of
the effect is to spectral flow the $SL(2)$ and $SU(2)$ components
of the vertex operator. More details on our spectral flow conventions
can be found in appendix B.
Moreover, the uncharged vertex operator $V_{2j+1}$ takes for $\OO^-_{2j+1}$
the form (see eq.\ \bdn)
\eqn\cac{
V_{2j+1}=\frac{k}{(2j+1)^2}\Phi^{(sl)}_{j,-j,-j}\Phi^{(su)}_{j,-j,-j}
~.}

As in the $\NN=2$ topological string theory \refs{\WittenZZ,\WarnerZH}, a correlation 
function in the $\NN=4$ topological string can be computed 
in the NS sector of the untwisted theory by inserting an extra spectral flow
operator with $\theta=1$ to saturate the background charge $-\hat c=-2$
induced by the topological twist. In our case, the appropriate insertion 
is the $\NN=4$ current $J^{--}$. For instance, the $3$-point function 
\caa\ can be computed in the untwisted theory as
\eqn\cad{
c_{j_1j_2j_3}=\Big\langle \Big ({J'}^{--}\bar {J'}^{--}\OO^-_{2j_1+1}\Big)
\OO^-_{2j_2+1}V_{2j_3+1}\Big\rangle
~,}
where we have defined the $U(1)$ gauge equivalent current ${J'}^{--}$ as
\eqn\cae{
{J'}^{--}=e^{-\int (J+J_g)}=J^{--}e^{-\int J_g}
~.}
As a result,
\eqn\caf{
{J'}^{--}\bar {J'}^{--}\OO^-_{2j_1+1}=
e^{iH_1+i\bar H_1}\Phi^{(sl)}_{j,-j-1,-j-1}\Phi^{(su)}_{j,-j,-j}
~}
and we see that the $H_1$, $\bar H_1$ momentum in
\cad\ is automatically conserved. Thus, \cad\ reduces to\foot{In this paper
we omit certain extra factors that depend explicitly on the
worldsheet variables. When taken into account
properly they cancel out at the end of the computation
and isolate the dimensionless part of the untwisted $N$-point functions. 
See \NakamuraSM\ for a careful treatment of a similar calculation in the 
$\NN=2$ topological string.}
\eqn\cag{\eqalign{
c_{j_1j_2j_3}=\frac{k}{(2j_3+1)^2}&
\Big\langle \Phi^{(sl)}_{j_1,-j_1-1,-j_1-1} \Phi^{(sl),w=1}_{j_2,-j_2-1,-j_2-1}
\Phi^{(sl)}_{j_3,-j_3,-j_3}\Big\rangle\times
\cr
&\Big\langle \Phi^{(su)}_{j_1,-j_1,-j_1} \Phi^{(su),w=1}_{j_2,-j_2,-j_2}
\Phi^{(su)}_{j_3,-j_3,-j_3}\Big\rangle
~.}}

This amplitude can be computed explicitly using known 
results for the three-point functions in the $SL(2)$ and $SU(2)$
WZW models \refs{\ZamolodchikovBD,\TeschnerFT,\TeschnerUG}. 
Since this is essentially the computation performed in \AharonyVK\
we will not repeat it here. Instead, we proceed to calculate \cag\   
with the use of the SRT map \refs{\RibaultWP,\RibaultMS}, 
which is summarized in appendix C. 
The fact that the SRT map can be very useful in the context of
the topological strings was pointed out for the first time in \NakamuraSM.
For the $SL(2)$ part in \cag\ we get a maximally winding number violating
amplitude, which can be recast into the form
\eqn\cai{\eqalign{
&\Big\langle \Phi^{(sl)}_{j_1,-j_1-1,-j_1-1} \Phi^{(sl),w=1}_{j_2,-j_2-1,-j_2-1}
\Phi^{(sl)}_{j_3,-j_3,-j_3}\Big\rangle=-\delta\bigg(j_1+j_2+j_3-\frac{k-2}{2}\bigg)
\cr
&\frac{2\pi^{-3}}{\sqrt k} c_{k+2} (2j_3+1)^2
\prod_{i=1}^3\gamma(-1-2j_i) 
\Big\langle \VV_{\frac{j_1+1}{\sqrt k}+\frac{\sqrt k}{2}}
\VV_{\frac{j_2+1}{\sqrt k}+\frac{\sqrt k}{2}}
\VV_{\frac{j_3+1}{\sqrt k}+\frac{\sqrt k}{2}} \Big\rangle_{Liouville}
~.}}
We are using the standard $\gamma$-function notation
\eqn\caj{
\gamma(x)=\frac{\Gamma(x)}{\Gamma(1-x)}
~.}
The three-point function on the r.h.s.\ is a three-point function in 
Liouville theory with linear dilaton slope $Q=\frac{1}{\sqrt k}+\sqrt k$.
The vertex operators $\VV_a$ are primary fields of this theory
with asymptotic (weak-coupling) form
\eqn\caja{
\VV_a=e^{\sqrt 2 a \phi}
~}
and scaling dimension $\Delta(\VV_a)=a(Q-a)$. $\phi$ is the Liouville field. 
In \cai\ $c_{k+2}$ is a (possibly diverging) $k$-dependent constant, 
which is part of the SRT map \RibaultMS.

With analytic continuation in $k$ ($i.e.$ $k\rightarrow -k$)
we obtain similar expressions for the $SU(2)$ part 
\eqn\cak{\eqalign{
&\Big\langle \Phi^{(su)}_{j_1,-j_1,-j_1} \Phi^{(su),w=1}_{j_2,-j_2,-j_2}
\Phi^{(su)}_{j_3,-j_3,-j_3}\Big\rangle=-\delta\bigg(j_1+j_2+j_3-\frac{k-2}{2}\bigg)
\cr
&i\frac{2\pi^{-3}}{\sqrt k}\tilde c_{k-2}\prod_{i=1}^3 \gamma(1+2j_i)
\Big\langle \widetilde \VV_{\frac{j_1}{\sqrt k}+\frac{\sqrt k}{2}}
\widetilde \VV_{\frac{j_2}{\sqrt k}+\frac{\sqrt k}{2}}
\widetilde \VV_{\frac{j_3}{\sqrt k}+\frac{\sqrt k}{2}} \Big\rangle_{Coulomb-Gas}
~.}}
The three-point function on the r.h.s.\ of this equation is a
three-point function in the Coulomb-Gas representation of the
$(1,k)$ minimal model. In this representation the vertex operators
$\tilde \VV_a$ are
\eqn\caka{
\tilde \VV_a=e^{i\sqrt 2 a \tilde \phi}
~,}
where $\tilde \phi$ is the boson of the Coulomb-Gas representation.
The scaling dimension of $\tilde \VV_a$ is $\Delta(\tilde \VV_a)=a(\tilde Q+a)$,
where $\tilde Q=\frac{1}{\sqrt k}-\sqrt k$. 

Combining the above expressions with the fact that\foot{For a 
summary of the basic properties of the $(1,k)$ minimal bosonic
string see appendix D.} 
\eqn\cala{
\TT_{2j+1}=c\bar c~ \VV_{\frac{j+1}{\sqrt k}+\frac{\sqrt k}{2}}
\widetilde \VV_{\frac{j}{\sqrt k}+\frac{\sqrt k}{2}}
}
is a physical tachyon vertex operator in the $(1,k)$ minimal bosonic string
we obtain
\eqn\cam{
c_{j_1j_2j_3}=-i \delta\bigg(j_1+j_2+j_3-\frac{k-2}{2}\bigg)
4\pi^{-6} c_{k+2}\tilde c_{k-2} \prod_{i=1}^3 \frac{1}{(1+2j_i)^2} 
\Big\langle \TT_{2j_1+1}\TT_{2j_2+1}\TT_{2j_3+1}\Big\rangle_{(1,k)}
~.}
We see that, up to a normalization factor, the $\NN=4$
topological string three-point function $c_{j_1j_2j_3}$
is the same as a corresponding three-point function in the $(1,k)$ minimal
bosonic string. In addition, \cam\ suggests a correspondence
between the $\NN=4$ topological string vertex operator $\OO^-_{2j+1}$
and the $(1,k)$ minimal bosonic string tachyon 
$\TT_{2j+1}$. Later in this section, we will test this identification
by computing higher $N$-point functions.

\subsec{4-point functions}

It is known \BerkovitsVY\ that the general four-point function in
the $\NN=4$ topological string takes the form
\eqn\cca{
c_{j_1j_2j_3j_4}=\bigg\langle \phi_{j_1}\phi_{j_2}
\bigg(J^{--}\bar J^{--}\phi_{j_3}\bigg)\int d^2 z_4 \phi_{j_4}\bigg\rangle
~.}
As in the previous subsection, here also we will focus on the correlation
functions of the vertex operators $\OO^-_{2j+1}$.
Again, we can compute this amplitude in the untwisted theory
by inserting an extra $J^{--}$ current to saturate the background
charge induced by the topological twist. 
The untwisted theory amplitude reads
\eqn\ccb{
c_{j_1j_2j_3j_4}=\bigg\langle 
\bigg({J'}^{--}\bar {J'}^{--}\OO^{-}_{2j_1+1}\bigg)\OO^-_{2j_2+1}
\bigg({J'}^{--}\bar {J'}^{--}\OO^{-}_{2j_3+1}\bigg)
\int d^2 z_4~ \OO^-_{2j_4+1}\bigg\rangle
~.}

Using the expressions \cab\ and \caf\ we see immediately that
the $H_1$, $\bar H_1$ momentum in \ccb\ is automatically conserved
and that the amplitude reduces to a product of $SL(2)$ and $SU(2)$
parts
\eqn\ccc{\eqalign{
c_{j_1j_2j_3j_4}=\int d^2 z_4 &\bigg\langle
\Phi^{(sl)}_{j_1,-j_1-1,-j_1-1}\Phi^{(sl),w=1}_{j_2,-j_2-1,-j_2-1}
\Phi^{(sl)}_{j_3,-j_3-1,-j_3-1} \Phi^{(sl),w=1}_{j_4,-j_4-1,-j_4-1}\bigg\rangle
\times
\cr
&\bigg\langle \Phi^{(su)}_{j_1,-j_1,-j_1}\Phi^{(su),w=1}_{j_2,-j_2,-j_2}
\Phi^{(su)}_{j_3,-j_3,-j_3} \Phi^{(su),w=1}_{j_4,-j_4,-j_4}
\bigg\rangle
~.}}
With the use of the SRT map we obtain the following results.
For the $SL(2)$ and $SU(2)$ parts respectively we find 
\eqn\ccd{\eqalign{
&\bigg\langle\Phi^{(sl)}_{j_1,-j_1-1,-j_1-1}\Phi^{(sl),w=1}_{j_2,-j_2-1,-j_2-1}
\Phi^{(sl)}_{j_3,-j_3-1,-j_3-1} \Phi^{(sl),w=1}_{j_4,-j_4-1,-j_4-1}\bigg\rangle=
\frac{2 c^2_{k+2} }{\pi^5\sqrt k} \times
\cr
&\delta\bigg(\sum_{i=1}^4j_i-(k-2)\bigg)
\prod_{i=1}^4\gamma(-1-2j_i)
\bigg\langle\VV_{\frac{j_1+1}{\sqrt k}+\frac{\sqrt k}{2}}
\VV_{\frac{j_2+1}{\sqrt k}+\frac{\sqrt k}{2}} \VV_{\frac{j_3+1}{\sqrt k}+\frac{\sqrt k}{2}}
\VV_{\frac{j_4+1}{\sqrt k}+\frac{\sqrt k}{2}}\bigg\rangle_{Liouville}
,}}
\eqn\cce{\eqalign{
&\bigg\langle \Phi^{(su)}_{j_1,-j_1,-j_1}\Phi^{(su),w=1}_{j_2,-j_2,-j_2}
\Phi^{(su)}_{j_3,-j_3,-j_3}\Phi^{(su),w=1}_{j_4,-j_4,-j_4}\bigg\rangle=
-\frac{2i~\tilde c^2_{k-2}}{\pi^5 \sqrt k}\prod_{i=1}^4 \gamma(1+2j_i)\times
\cr
&\delta\bigg(\sum_{i=1}^4j_i-(k-2)\bigg)
\bigg\langle\tilde\VV_{\frac{j_1}{\sqrt k}+\frac{\sqrt k}{2}}
\tilde \VV_{\frac{j_2}{\sqrt k}+\frac{\sqrt k}{2}}
\tilde \VV_{\frac{j_3}{\sqrt k}+\frac{\sqrt k}{2}}
\tilde \VV_{\frac{j_4}{\sqrt k}+\frac{\sqrt k}{2}}\bigg\rangle_{Coulomb-Gas}
~.}}
From \ccc, \ccd\ and \cce\ we deduce that the 
topological string four-point function $c_{j_1j_2j_3j_4}$ can be written as
\eqn\ccf{\eqalign{
c_{j_1j_2j_3j_4}=&-\frac{4i ~ c^2_{k+2}\tilde c^2_{k-2}}{\pi^{10} k}
\delta\bigg(\sum_{i=1}^4j_i-(k-2)\bigg)
\prod_{i=1}^4 \frac{1}{(1+2j_i)^2}\times
\cr
&\bigg\langle \TT_{2j_1+1}\TT_{2j_2+1}\TT_{2j_3+1} \int d^2 z_4 ~\TT_{2j_4+1}
\bigg\rangle_{(1,k)}
~.}}
This is the four-point function generalization of eq.\ \cam\ above.
As before, we find that, up to a normalization factor that can be absorbed into the definition 
of the $\NN=4$ topological string vertex operators, the four-point functions
$c_{j_1j_2j_3j_4}$ are captured by corresponding four-point functions
in the $(1,k)$ minimal bosonic string. 

In the minimal bosonic string it is known independently \WittenMK\ (see also \DijkgraafDJ)
that
\eqn\ccg{\eqalign{
&\delta\bigg(\sum_{i=1}^4 j_i -(k-1)\bigg)
\bigg\langle \TT_{2j_1+1}\TT_{2j_2+1}\TT_{2j_3+1} \int d^2 z_4 ~\TT_{2j_4+1}
\bigg\rangle_{(1,k)} =
\cr 
&\delta\bigg(\sum_{i=1}^4j_i-(k-1)\bigg)\min(2j_i,k-2j_i-1)
~.}}
Unfortunately, we cannot make use of this result to determine the amplitude
appearing in the r.h.s.\ of \ccf, because the selection rule is different.
On the other hand, from eq.\ (6.8) in \AharonyVK\ we expect the 
$\NN=4$ topological string amplitude
\eqn\cci{
c_{j_1j_2j_3j_4}=\delta\bigg(\sum_{i=1}^4j_i-(k-2)\bigg)\min(2j_i+1,k-2j_i-1)
~,}
where an unimportant normalization factor has been dropped.
This result was anticipated in \AharonyVK\
as a non-trivial consistency requirement for heterotic-type II string duality.
Here we find it to be equivalent (up to a normalization factor) to the following 
statement in the minimal bosonic string 
\eqn\ccj{\eqalign{
&\delta\bigg(\sum_{i=1}^4 j_i -(k-2)\bigg)
\bigg\langle \TT_{2j_1+1}\TT_{2j_2+1}\TT_{2j_3+1} \int d^2 z_4 ~\TT_{2j_4+1}
\bigg\rangle_{(1,k)} =
\cr 
&\delta\bigg(\sum_{i=1}^4j_i-(k-2)\bigg)\min(2j_i+1,k-2j_i-1)
~.}}
We leave an explicit derivation of this result to a
future publication. Moreover, it would be interesting to obtain a better
understanding of the r\^ole of the above selection rules in the emerging relation 
between the minimal bosonic string and the minimal $\NN=4$
topological string.

\subsec{$N$-point functions}

For $N\geq 5$ the $N$-point amplitudes of the $\NN=4$ topological
string take the form ($c.f.$ \BerkovitsVY)
\eqn\cda{\eqalign{
c_{j_1j_2...j_N}=&
\bigg\langle \bigg({J'}^{--}\bar {J'}^{--}\OO^-_{2j_1+1}\bigg)
\OO^-_{2j_2+1}\bigg({J'}^{--}\bar {J'}^{--}\OO^-_{2j_3+1}\bigg)\times
\cr
&\int d^2 z_4~ \OO^-_{2j_4+1} 
\prod_{\ell=5}^N \int d^2 z_{\ell}~ \widehat G^-\widehat{\bar G}^+ 
\cdot \OO^-_{2j_{\ell}+1}\bigg\rangle
~.}}
The generators $\widehat G^-$, $\widehat{\bar G}^+$
 are defined as
\eqn\cdb{
\widehat G^-=u_1G^- +u_2\tilde G^-~, ~ ~
\widehat{\bar G}^+= \bar u_1 \bar G^++\bar u_2\bar{\tilde G}^+
~,}
where $u_1, u_2$ and $\bar u_1, \bar u_2$ are two independent sets of
complex numbers with the property 
\eqn\cdc{
|u_1|^2+|u_2|^2=1~,  ~ ~ |\bar u_1|^2+|\bar u_2|^2=1
~.}
As a result, the amplitude \cda\ is a homogeneous 
polynomial of the parameters 
$u_1,u_2,\bar u_1,\bar u_2$ of degree $2(N-4)$ . 
The polynomial expansion can be written as
\eqn\cdd{
c_{j_1j_2...j_N}(u_1,u_2;\bar u_1,\bar u_2)=
\sum_{n=0}^{N-4}
\CC^{n}_{j_1j_2...j_N}~ (u_1\bar u_1)^n 
(u_2\bar u_2)^{N-4-n}
~,}
where $\CC^{n}_{j_1j_2...j_N}$ are non-trivial data
of the $\NN=4$ topological string that can be computed by 
calculating \cda. 

This can be achieved by applying the SRT
map as before. First we need to determine the explicit
form of $\widehat G^-\widehat{\bar G}^+ 
\cdot \OO^-_{2j+1}$. By straightforward algebra 
and the use of the identity 
$\Phi^{(su),w=1}_{j,-j}=\Phi^{(su)}_{\frac{k-2}{2}-j,\frac{k-2}{2}-j}$
we find
\eqn\cdda{
G^-\cdot \OO^-_{2j+1}=
\frac{1}{\sqrt k}\Big[ (-2j-2)\Phi^{(sl),w=1}_{j,-j-2}\Phi^{(su),w=1}_{j,-j}+
(k-2-2j) e^{-iH_1+iH_2}\Phi^{(sl),w=1}_{j,-j-1}\Phi^{(su)}_{\frac{k-2}{2}-j,\frac{k-2}{2}-j-1}
\Big]
~.}
In the BRST cohomology of the coset, this is equivalent to
\eqn\cddb{
G^-\cdot \OO^-_{2j+1}=
\frac{1}{\sqrt k}\Big[ (-2j-2)\Phi^{(sl),w=1}_{j,-j-2} \Phi^{(su),w=1}_{j,-j} +
(k-2-2j) \Phi^{(sl)}_{j,-j-1}\Phi^{(su),w=-1}_{\frac{k-2}{2}-j,\frac{k-2}{2}-j-1}
\Big]
~.}
In a similar fashion,
\eqn\cddc{
\tilde G^-\cdot \OO^-_{2j+1}=
\frac{1}{\sqrt k}\Big[ (k-2j)~\Phi^{(sl),w=-1}_{-j+\frac{k-2}{2},-j+\frac{k}{2}+1}\Phi^{(su)}_{j,-j} +
2j~ \Phi^{(sl),w=1}_{j,-j-1}\Phi^{(su),w=1}_{j,-j+1}
\Big]
~.}
The action of $\widehat{\bar G}^+$ on the right-movers is
identical.

It is clear from the above expressions that the general 
coefficient $\CC^n_{j_1j_2...j_N}$ will not reduce to a
product of $SL(2)$ and $SU(2)$ amplitudes with maximal
winding number violation. In fact, the winding number violation
and the associated selection rules will depend crucially 
on how many times we insert the second term of \cddb\
or the first term of \cddc\ into the amplitude \cda.\foot{I would
like to thank D. Sahakyan and T. Takayanagi for a useful discussion
on this point.}
In order to illustrate the computation of the general coefficient
$\CC^n_{j_1j_2...j_N}$ let us consider in detail a few 
representative cases.

Consider, for example, the case of the amplitude
$\CC^{N-4}_{j_1j_2...j_N}$. It takes the form
\eqn\cdi{\eqalign{
\CC^{N-4}_{j_1j_2...j_N}=&
\bigg\langle \bigg({J'}^{--}\bar {J'}^{--}\OO^-_{2j_1+1}\bigg)
\OO^-_{2j_2+1}\bigg({J'}^{--}\bar {J'}^{--}\OO^-_{2j_3+1}\bigg)\times
\cr
&\int d^2 z_4~ \OO^-_{2j_4+1} 
\prod_{\ell=5}^N \int d^2 z_{\ell}~ G^- \bar G^+ 
\cdot \OO^-_{2j_{\ell}+1}\bigg\rangle
~.}}
We can choose the selection rule appropriately, so
that \cdi\ does not receive any contributions from insertions
of $G^{(su)-}$ (see the second term in \cddb).
In that case, \cdi\ reduces to the following product of 
maximally winding number violating $SL(2)$ and 
$SU(2)$ amplitudes
\eqn\cdj{\eqalign{
&\CC^{N-4}_{j_1j_2...j_N}=\frac{1}{k^{N-4}}\int d^2 z_4
\prod_{\ell=4}^N\int d^2 z_{\ell} ~(2j_{\ell}+2)^2
\cr
&\bigg\langle\Phi^{(sl)}_{j_1,-j_1-1,-j_1-1}\Phi^{(sl),w=1}_{j_2,-j_2-1,-j_2-1}
\Phi^{(sl)}_{j_3,-j_3-1,-j_3-1}
\Phi^{(sl),w=1}_{j_4,-j_4-1,-j_4-1} 
\Phi^{(sl),w=1}_{j_{\ell},-j_{\ell}-2,-j_{\ell}-2}\bigg\rangle \times
\cr
&\bigg\langle \Phi^{(su)}_{j_1,-j_1,-j_1}\Phi^{(su),w=1}_{j_2,-j_2,-j_2}
\Phi^{(su)}_{j_3,-j_3,-j_3}\Phi^{(su),w=1}_{j_4,-j_4,-j_4}
\Phi^{(su),w=1}_{j_{\ell},-j_{\ell},-j_{\ell}}
\bigg\rangle
~.}}
The $H_1$ and $\bar H_1$
momenta are automatically conserved as always.
Then by direct application of the SRT map we find again that we can
recast the $\NN=4$ topological string amplitude $\CC^{N-4}_{j_1j_2...j_N}$
in terms of an amplitude in the $(1,k)$ minimal bosonic string
\eqn\cdk{\eqalign{
\CC^{N-4}_{j_1j_2...j_N}=&\frac{-4i \pi^{6-4N}}{k^{N-3}}
(c_{k+2}\tilde c_{k-2})^{N-2}\prod_{i=1}^N \frac{1}{(1+2j_i)^2}
\delta\bigg(\sum_{i=1}^N j_i-\frac{k-2}{2}(N-2)\bigg)\times
\cr
&\bigg\langle \TT_{2j_1+1}\TT_{2j_2+1}\TT_{2j_3+1} 
\prod_{\ell=4}^N \int d^2 z_{\ell} ~\TT_{2j_{\ell}+1}\bigg \rangle_{(1,k)}
~.}}

This computation generalizes to all the coefficients
$\CC^n_{j_1j_2...j_N}$ (for $n=0,...,N-4$)
without any $e^{iH_2}K^-$ or $e^{iH_2} J^+$
insertions. In that case, the general result is
\eqn\cdl{\eqalign{
\CC^{n}_{j_1j_2...j_N}=&\frac{-4i \pi^{6-4N}}{k^{N-3}}
(c_{k+2}\tilde c_{k-2})^{N-2}\prod_{i=1}^N \frac{1}{(1+2j_i)^2}
\delta\bigg(\sum_{i=1}^N j_i-\frac{k-2}{2}(N-2)-n\bigg)\times
\cr
&\bigg\langle \TT_{2j_1+1}\TT_{2j_2+1}\TT_{2j_3+1} 
\prod_{\ell=4}^N \int d^2 z_{\ell} ~\TT_{2j_{\ell}+1}\bigg \rangle_{(1,k)}
~.}}

There are, however, many other cases, with a non-zero number
of $e^{iH_2}K^-$ or $e^{iH_2} J^+$ insertions that have different 
selection rules and do not reduce to maximally winding number violating
amplitudes. Consider, for example, the amplitude 
\cdi\ with selection rules $\sum_{i=1}^N j_i =\frac{k-2}{2}(N-2-\ell)-\ell$
for $\ell=1,2,...,N-4$. In this case, \cdi\ will receive contributions from 
those terms that have $N-4-\ell$ $G^{(sl)-}=e^{iH_1}J^-$ insertions
and $\ell$ $G^{(su)-}=e^{iH_2}K^-$ insertions. 
Since they do not exhibit
maximal winding number violation these correlation functions
do not appear to have a clear interpretation in 
the minimal bosonic string. After the application of the SRT map
the amplitudes will involve extra integrated insertions of degenerate
vertex operators. In particular, this makes the identification 
of vertex operators $\OO^-_{2j+1}\leftrightarrow \TT_{2j+1}$
less straightforward. Similar statements apply
to $\CC^n_{j_1j_2...j_N}$ for any $n=0,1,...,N-4$.
It is important to understand this point further.

\newsec{A web of equivalences}

In this paper we identified a certain class of physical states in the
minimal $\NN=4$ topological string and proceeded to analyze their $N$-point 
functions for arbitrary $N$. With the use of the
SRT map, we found that we can recast a subset of these correlation functions 
in terms of corresponding amplitudes in the 
$(1,k)$ minimal bosonic string. The results presented in 
this paper have immediate consequences for the topological sector of little string theory. 
With the use of the minimal bosonic string we can, in principle, deduce
in little string theory the precise value of a large number of correlation functions of
off-shell observables. For example, we can recast 
a result anticipated in \AharonyVK\ as a non-trivial requirement of heterotic-type II string
duality (see eq.\ \cci) as a statement in the minimal bosonic string, which we expect
can be verified explicitly. It would be interesting to explore this point further
with an explicit computation of the minimal bosonic string amplitudes
and see how much information
we can obtain about the structure of little string theory from the
correlation functions presented in this paper. We hope to return to
this issue in a future publication.

More generally, the above results suggest an interesting relation
between the $\NN=4$ topological string and the $(1,k)$ minimal 
bosonic string.\foot{Earlier indications of such a relation 
can be found in \refs{\KonechnyDF, \TakayanagiYJ} (see also \Rastelli).} 
It is natural to ask whether this relation is a full-fledged
equivalence like the equivalence between the $\NN=2$
topological string on $SL(2)_1/U(1)$ and the $c=1$ non-critical
bosonic string at self-dual radius \MukhiZB. For this it is
crucial to obtain a better understanding of the  
amplitudes \cda\ with general selection rules and to examine whether
a similar relation persists for amplitudes that involve generic
states in the cohomology of the minimal $\NN=4$ topological string.
The latter requires a full classification of the observables of the minimal $\NN=4$ 
topological string, which is beyond the immediate scope of this note.

Clearly, a full equivalence would be a powerful statement
with many interesting implications. For instance, when combined
with other known facts, it would imply
that the $(1,k)$ minimal bosonic string is equivalent to the
following seemingly unrelated theories:
\item{(1)} The minimal $\NN=4$ topological string, which is also 
related via T-duality to the topological string on K3 near an
ADE singularity and via holography to the topological
sector of little string theory \refs{\OoguriWJ,\GiveonZM,\GiveonPX}.
\item{(2)} The double-scaled ADE matrix models of \KostovIE.
\item{(3)} The generalized Kontsevich matrix models 
\refs{\KontsevichTI,\AganagicQJ}.
\item{(4)} The $\NN=2$ topologically twisted coset $SU(2)_k/U(1)$
coupled to topological gravity 
\refs{\EguchiVZ\DijkgraafNC\DijkgraafDJ\LiKE\WittenMK-\DijkgraafQH}.
\item{(5)} The B-model $\NN=2$ topological string on a hypersurface of the form
\AganagicQJ
\eqn\daa{
z_1^k+z_2+z_3^2+z_4^2=0
~.}
\item{(6)} The B-model $\NN=2$ topological string on $H_3^+\times S^3$ 
\RastelliPH.

\noindent
A better understanding of the structure of the minimal $\NN=4$ 
topological string both in the closed string and open string sector
would determine whether (1) is a bona fide member of this set
of theories.

\vskip10mm
\centerline{\bf Acknowledgments}
\vskip0.2cm

I would like to thank D. Sahakyan and T. Takayanagi for a discussion
and N. Prezas for a useful comment.

\appendix{A}{The $\NN=4$ superconformal algebra}

The (small) $\NN=4$ supeconformal algebra (SCA)
comprises of the $\NN=2$ superconformal algebra generators $(T,G^{\pm},J)$
plus two additional currents of charge $\pm 2$, which
will be denoted as $J^{++}$ and $J^{--}$.\foot{We are following the notation of
\BerkovitsVY.} 
The three currents $(J,J^{++},J^{--})$ form an $SU(2)$ algebra,
under which the $G^{\pm}$ currents can be used to generate
two more fermionic generators $\tilde G^{\pm}$.
Altogether these four fermionic generators form
the $SU(2)$ doublets
\eqn\brstgaa{
(G^+,\tilde G^-)~, ~ ~ (\tilde G^+, G^-)
~.}
In unitary theories the adjoint is defined by
\eqn\brstgab{
G^{+\dagger}=G^- ~, ~ ~ \tilde G^{+\dagger}=\tilde G^- ~,
~ ~ J^{++\dagger}=J^{--}
~.}

The $\NN=4$ SCA OPE's are
\eqn\brstgac{
J^{--}(z) J^{++}(0)\sim \frac{\hat c}{2z^2}-\frac{J(0)}{z}
~,}
\eqn\brstgad{
J^{--}(z)G^{+}(0)\sim \frac{\tilde G^-(0)}{z}~, ~ ~
J^{++}(z)\tilde G^-(0) \sim -\frac{G^+(0)}{z}
~,}
\eqn\brstgae{
J^{++}(z)G^{-}(0)\sim \frac{\tilde G^+(0)}{z}~, ~ ~
J^{--}(z)\tilde G^+(0)\sim -\frac{G^-(0)}{z}
~,}
\eqn\brstgaea{
J(z)G^{\pm}(0)\sim \pm \frac{G^{\pm}(0)}{z}~, ~ ~
J(z)\tilde G^{\pm}(0)\sim \pm \frac{\tilde G^{\pm}(0)}{z} 
~,}
\eqn\brstgaf{
J^{--} \cdot G^- \sim J^{++}\cdot G^+ \sim J^{++} \cdot \tilde G^+
\sim J^{--} \cdot \tilde G^-\sim0
~,}
\eqn\brstgag{
G^{\pm}\cdot G^{\pm}\sim \tilde G^{\pm}\cdot \tilde G^{\pm}\sim
G^+\cdot \tilde G^-\sim G^- \cdot \tilde G^+\sim 0
~,}
\eqn\brstgai{
G^+(z)G^-(0)\sim \frac{\hat c}{z^3}+\frac{J(0)}{z^2}+\frac{2T(0)+\d J(0)}{z}
~,}
\eqn\brstgaj{
\tilde G^+(z)\tilde G^-(0)\sim \frac{\hat c}{z^3}+\frac{J(0)}{z^2}+\frac{2T(0)+\d J(0)}{z}
~,}
\eqn\brstgak{
G^+(z)\tilde G^+(0)\sim \frac{J^{++}(0)}{z^2}+\frac{\d J^{++}(0)}{2z}
~,}
\eqn\brstgal{
G^-(z)\tilde G^-(0)\sim \frac{J^{--}(0)}{z^2}+\frac{\d J^{--}(0)}{2z}
~.}

\noindent
Any theory with $\hat c=2$, $\NN=2$ superconformal symmetry and 
integral $U(1)$ charges automatically has $\NN=4$ superconformal symmetry 
with the extra generators provided by the spectral flow operators 
$J^{\pm\pm}=e^{\pm \int J}$.

\appendix{B}{$SL(2,\IR)$ and $SU(2)$ conventions}

In this appendix, we summarize our conventions for the
$SL(2)$ and $SU(2)$ WZW models. For $SL(2)$
we follow the conventions of \RibaultWP, where
\eqn\appda{
\Phi^{(sl)}_{j,m,\bar m}=\int d^2x~ x^{-j-1+m} \bar x^{-j-1+\bar m} 
\Phi^{(sl)}_j(x,\bar x)
~}
and
\eqn\appdaa{
J^3(z)\Phi^{(sl)}_{j,m}(0)\sim \frac{m}{z}\Phi^{(sl)}_{j,m}(0)~, ~ ~
J^{\pm}(z)\Phi^{(sl)}_{j,m}(0)\sim \frac{m\pm (j+1)}{z}\Phi^{(sl)}_{j,m\pm 1}(0)
~.}
The semiclassical behavior of $\Phi^{(sl)}(x)$ in the $(\gamma,\bar \gamma, \phi)$
coordinate system is \RibaultWP
\eqn\appdb{
\Phi^{(sl)}_{j,cl}(x)=\frac{2j+1}{\pi}\Big(|\gamma-x|^2e^{\phi}+e^{-\phi}\Big)^{2j}
~.}
At the conformal boundary of $SL(2)$ $\phi\rightarrow \infty$ and
$\Phi^{(sl)}_{j,cl}$ can be expanded as (see $e.g.$ eq.\ (2.8) in \KutasovXU)
\eqn\appdc{
\Phi^{(sl)}_{j,cl}\simeq -e^{-2(j+1)\phi}\delta^{(2)}(\gamma-x)+
\OO(e^{-2(j+2)\phi})-\frac{(2j+1)|\gamma-x|^{4j}}{\pi}e^{2j\phi}
+\OO(e^{2(j-1)\phi})
~.}
As a result, for $j>0$, $\Phi^{(sl)}_{j,m,\bar m}$ will give rise to
a normalizable vertex operator in \baa\
and $\Phi^{(sl)}_{-j-1,m,\bar m}$ will give rise to non-normalizable one. 
The two vertex operators are related by the reflection relation
\eqn\appdca{
\Phi^{(sl)}_{j,m,\bar m}=R(j,m,\bar m)\Phi^{(sl)}_{-j-1,m,\bar m}
~,}
where
\eqn\appdcb{
R(j,m,\bar m)=k\bigg[\frac{1}{k\pi}\gamma\bigg(\frac{1}{k}\bigg)\bigg]^{-2j-1}
\gamma\bigg(1+\frac{2j+1}{k}\bigg) 
\frac{\Gamma(2j+1)\Gamma(-j-m)\Gamma(-j+\bar m)}
{\Gamma(-2j)\Gamma(1+j-m)\Gamma(1+j+\bar m)}
~}
denotes the $SL(2,\IR)$ reflection coefficient.

In $SU(2)$ 
the OPE's of the currents with the primary fields $\Phi^{(su)}_{j,m}$ are
\eqn\appdda{
K^3(z)\Phi^{(su)}_{j,m}(0)\sim \frac{m}{z}\Phi^{(su)}_{j,m}(0)~, ~ ~
K^{\pm}(z)\Phi^{(su)}_{j,m}(0)\sim \frac{j\mp m}{z}\Phi^{(su)}_{j,m\pm 1}(0)
~.}

For quick reference, we also summarize in this appendix our conventions
for the spectral flow operation in the $\NN=2$ superconformal algebra, the bosonic
$SL(2)_{k+2}$ WZW model and the bosonic $SU(2)_{k-2}$ WZW model.

For the $\NN=2$ superconformal algebra with central charge $c$ the
spectral flow transformation by an amount $\theta$ is an automorphism of the form
\eqn\appaa{
\tilde L_n=L_n+\theta J_n+\theta^2\frac{c}{6}\delta_{n,0}~, ~~
\tilde G^{\pm}_n=G^{\pm}_{n\pm \theta}~, ~ ~
\tilde J_n=J_n+\frac{c}{3}\theta
~.}
On the level of primary fields this transformation can be achieved
by multiplying with the vertex operator $e^{-i\theta\sqrt{\frac{c}{3}}X_R}$,
where $X_R$ is a canonically normalized boson that bosonizes
the $U(1)_R$ current.

For the $SL(2)_{k+2}$ WZW model the spectral flow operation with winding
number $w\in \IZ$ is defined by the current automorphism
\eqn\appab{
\tilde L_n=L_n-w J^3_n-\frac{k+2}{4} w^2\delta_{n,0}~, ~ ~
\tilde J^3_n=J^3_n-\frac{k+2}{2}w\delta_{n,0}~, ~ ~\tilde J_n^{\pm}=J^{\pm}_{n\mp w}
~.}
On vertex operators it corresponds to multiplication with the 
operator $e^{w\sqrt{\frac{k+2}{2}}X^{(sl)}}$. $X^{(sl)}$ is the canonically
normalized boson that bosonizes the current $J^3$, $i.e.$
\eqn\appac{
J^3=-\sqrt{\frac{k+2}{2}}\d X^{(sl)}
~.}
In a similar fashion, for the $SU(2)_{k-2}$ WZW model the spectral flow
operation with winding number $w\in \IZ$ is defined by the current
automorphism
\eqn\appad{
\tilde L_n=L_n+w K^3_n+\frac{k-2}{4} w^2\delta_{n,0}~, ~ ~
\tilde K^3_n=K^3_n-\frac{k-2}{2}w\delta_{n,0}~, ~ ~\tilde K_n^{\pm}=K^{\pm}_{n\pm w}
~.}
On vertex operators it corresponds to multiplication with the operator
$e^{iw\sqrt{\frac{k-2}{2}}X^{(su)}}$. $X^{(su)}$ is the canonically normalized boson
that bosonizes the current $K^3$, $i.e.$
\eqn\appae{
K^3=i\sqrt{\frac{k-2}{2}}X^{(su)}
~.}
The spectral flowed primary vertex operators $\Phi^{(sl)}_{j,m}$, 
$\Phi^{(su)}_{j,m}$ will be denoted as $\Phi^{(sl),w}_{j,m}$, $\Phi^{(su),w}_{j,m}$.

In the main text we obtain the $\frac{SL(2)_k}{U(1)}\times\frac{SU(2)_k}{U(1)}$
theory by gauging the null $U(1)$ current 
\eqn\appak{
J_g=i\d H_2-i\d H_1+K_3-J_3
~}
in $SL(2)_k\times SU(2)_k$. As in \refs{\TakayanagiYB, \NakamuraSM} 
it will be useful to define the gauge equivalent vertex operators 
\eqn\appal{
\OO\sim \OO ~e^{\pm \int J_g}=\OO ~e^{\pm (iH_2-iH_1+i\sqrt{\frac{k-2}{2}}X^{su}+
\sqrt{\frac{k+2}{2}}X^{sl})}
~.}
In particular, for $\OO=\OO^+_{2j+1}$ we get
\eqn\appam{
\OO^+_{2j+1}\sim e^{-\int J_g}~ \OO^+_{2j+1}=e^{-iH_2}\Phi^{(sl),w=-1}_{j,j+1}
\Phi^{(su),w=-1}_{j,j}
~.}
Using the $SL(2)$ representation equivalence 
$\DD^-_{-j-\frac{k+2}{2}}=\DD^{+,w=-1}_{j}$, $i.e.$
\eqn\appan{
\Phi^{(sl)}_{j+\frac{k}{2},-j-\frac{k+2}{2}}=\Phi^{(sl),w=-1}_{-j-1,-j}
}
and the corresponding $SU(2)$ identity
\eqn\appaga{
\Phi^{(su),w=-1}_{j,j}=\Phi^{(su)}_{\frac{k-2}{2}-j,j-\frac{k-2}{2}}
~}
we find
\eqn\appao{
\OO^+_{2j+1}= e^{-iH_2}\Phi^{(sl)}_{\frac{k-2}{2}-j,j-\frac{k}{2}}
\Phi^{(su)}_{\frac{k-2}{2}-j,j-\frac{k-2}{2}} 
= \OO^-_{k-2j-1}
~}
thus reproducing in the $\NN=4$ topological string 
the $\NN=2$ string relation that appears in eq.\ (5.34) of \AharonyVK. 
For reference, we also
list the representation identities 
\eqn\appaoa{
\Phi^{(sl),w=1}_{-j-1,j}=\Phi^{(sl)}_{j+\frac{k}{2},j+\frac{k+2}{2}}~, ~ ~
\Phi^{(su),w=1}_{j,-j}=\Phi^{(su)}_{\frac{k-2}{2}-j,\frac{k-2}{2}-j}
~.}

\appendix{C}{Summary of the Stoyanovsky-Ribault-Teschner map}

Recently the authors of \refs{\RibaultWP,\RibaultMS} formulated a
precise map between sphere correlation functions in the $SL(2,\IC)/SU(2)$
WZW model and correlation functions in Liouville field theory.
The basic formula that was obtained in \RibaultWP\ provides
a map between winding number conserving $N$-point functions in 
the $SL(2,\IC)/SU(2)$ WZW model at level $k$ and $(2N-2)$-point functions
in Liouville field theory with linear dilaton slope $Q=b+b^{-1}$ and 
$b^2=\frac{1}{k-2}$. In this paper, we are interested, as in \NakamuraSM,
in winding number violating amplitudes. 
For such amplitudes the relevant formula
is \RibaultMS\foot{Compared to the conventions of \RibaultMS\ we have
$m_{ours}=-m_{there}$, $\bar m_{ours}=-\bar m_{there}$ and
$w_{ours}=-w_{there}$.}
\eqn\appba{\eqalign{
&\bigg\langle \prod^{N}_{i=1}\Phi^{(sl),w_i}_{j_i,m_i,\bar m_i}(z_i,\bar z_i)\bigg\rangle
_{\sum w_i=r\geq 0}=
\prod^N_{i=1}\NN^{j_i}_{m_i,\bar m_i}\delta^{(2)}\bigg(\sum^N_{\ell=1}
m_{\ell}+\frac{k}{2}r\bigg)\times
\cr
&\prod^{N-2-r}_{a=1}\int d^2y_a~\tilde{\FF}_k(z_i,\bar z_i;y_a,\bar y_a)
\bigg\langle\VV_{\alpha_i}(z_i,\bar z_i)\VV_{-\frac{1}{2b}}(y_a,\bar y_a)\bigg\rangle
~,}}
where for $c_k$ a $k$-dependent constant,
\eqn\appbb{\eqalign{
\tilde \FF_k(z_i,\bar z_i;y_a,\bar y_a)&=\frac{2\pi^{3-2N}bc_k^r}{(N-2-r)!}
\prod_{i<j\leq N}(z_i-z_j)^{\beta_{ij}}(\bar z_i-\bar z_j)^{\bar \beta_{ij}}
\cr
&\prod_{a<b\leq N-2-r}|y_a-y_b|^k \prod^N_{r=1} \prod^{N-2-r}_{c=1}
(z_r-y_c)^{-m_r-\frac{k}{2}}(\bar z_r-\bar y_c)^{-\bar m_r-\frac{k}{2}}
~,}}
with
\eqn\appbc{\eqalign{
\beta_{ij}&=\frac{k}{2}+m_i+m_j-\frac{k}{2}w_i w_j-w_i m_j-w_j m_i~,
\cr
\bar\beta_{ij}&=\frac{k}{2}+\bar m_i+\bar m_j-\frac{k}{2}\bar w_i \bar w_j-\bar w_i\bar m_j
-\bar w_j\bar m_i
~}}
and
\eqn\appbca{
\NN^j_{m,\bar m}=\frac{\Gamma(-j+m)}{\Gamma(1+j-\bar m)}
~,}
\eqn\appbcb{
\alpha_i=bj_i+b+\frac{1}{2b}
~.}
$\VV_a$ are vertex operators in Liouville field theory with
asymptotic (weak coupling) form
\eqn\appbd{
\VV_a=e^{\sqrt 2 a \phi}
~.}
In our conventions $\alpha'=2$ and the Liouville interaction
takes the form $\mu_L e^{\sqrt 2 b\phi}$. The Liouville field theory
central charge is 
\eqn\appbe{
c_L=1+6Q^2
~.}
The scaling dimension of $\VV_a$ is
\eqn\appbf{
\Delta(\VV_a)=a(Q_L-a)
~.}

It will be useful to consider the analytic continuation of the above
expressions to $SU(2)_k$. This entails taking $k\rightarrow -k$ and setting
$j\rightarrow -j-1$, $m\rightarrow -m, \bar m\rightarrow - \bar m$. Then we find
that we can recast a correlation function of $SU(2)_k$ operators 
in terms of a (Coulomb-Gas) correlation function in a linear dilaton theory
with imaginary slope $\tilde b+\tilde b^{-1}$, 
where $\tilde b^2=-\frac{1}{k+2}$. The analytically continued version
of \appba\ reads
\eqn\appbg{\eqalign{
&\bigg\langle \prod^{N}_{i=1}\Phi^{(su),w_i}_{j_i,m_i,\bar m_i}(z_i,\bar z_i)\bigg\rangle
_{\sum w_i=r\geq 0}=
\prod^N_{i=1}\widetilde \NN^{j_i}_{m_i,\bar m_i}\delta^{(2)}\bigg(\sum^N_{\ell=1}
m_{\ell}+\frac{k}{2}r\bigg)\times
\cr
&\prod^{N-2-r}_{a=1}\int d^2y_a~\tilde{\tilde{\FF}}_k(z_i,\bar z_i;y_a,\bar y_a)
\bigg\langle\tilde \VV_{\alpha_i}(z_i,\bar z_i)\tilde \VV_{-\frac{1}{2\tilde b}}(y_a,\bar y_a)\bigg\rangle
~,}}
where for $\tilde c_k$ a $k$-dependent constant,
\eqn\appbi{\eqalign{
\tilde{\tilde \FF}_k(z_i,\bar z_i;y_a,\bar y_a)&=\frac{2\pi^{3-2N}\tilde b \tilde c_k^r}{(N-2-r)!}
\prod_{i<j\leq N}(z_i-z_j)^{-\beta_{ij}}(\bar z_i-\bar z_j)^{-\bar \beta_{ij}}
\cr
&\prod_{a<b\leq N-2-r}|y_a-y_b|^{-k} \prod^N_{r=1} \prod^{N-2-r}_{c=1}
(z_r-y_c)^{m_r+\frac{k}{2}}(\bar z_r-\bar y_c)^{\bar m_r+\frac{k}{2}}
~,}}
with $\beta_{ij}$ and $\bar \beta_{ij}$ as in \appbc\
and
\eqn\appbck{
\widetilde \NN^j_{m,\bar m}=\frac{\Gamma(1+j-m)}{\Gamma(-j+\bar m)}
~,}
\eqn\appbcl{
\alpha=-\tilde bj+\frac{1}{2\tilde b}
~.}
$\tilde \VV_a$ are vertex operators of the form $e^{\sqrt 2 a \tilde \phi}$
in the Coulomb-Gas representation
of the $(1, k+2)$ minimal model.

\appendix{D}{The $(1,k)$ minimal bosonic string}

The $(1,k)$ minimal bosonic string is obtained by 
coupling the $(1,k)$ minimal model with Liouville field theory
in such a way that the total central charge is 26.
Despite the fact that the $(1,k)$ minimal models are outside
the range of definition of the ``minimal'' Virasoro series the $(1,k)$ 
minimal bosonic string is well-defined \DijkgraafQH.
The minimal model has central charge
\eqn\appea{
c=1-6\frac{(k-1)^2}{k}
~.}
This implies that the linear dilaton slope of Liouville
theory is $Q=\frac{1}{\sqrt k}+\sqrt k$.

The $(1,k)$ minimal string has an infinite number of physical states $T_{r,s}$,
which are labeled by two integers $r,s$ \refs{\BershadskyPE,\RastelliPH}. 
The corresponding vertex operators are
\eqn\appeb{
T_{r,s}=c W_{r,s}e^{\sqrt 2 \big(\frac{r+1}{2\sqrt k}+\frac{s}{2}\sqrt k\big)\phi}~,  ~ ~
r=1,2,...,k-1~, ~ s=1,2,3,...
~,}
where $\phi$ is the Liouville field theory boson and $W_{r,s}$
are states of the $(1,k)$ minimal model. In the Coulomb-Gas representation
the latter take the form 
\eqn\appec{
W_{r,s}=e^{i\sqrt 2 \big(\frac{r-1}{2\sqrt k}+\frac{s}{2}\sqrt k\big)\tilde \phi}
~.}
$\tilde \phi$ is a free boson with linear dilaton slope
$\tilde Q=\frac{1-k}{\sqrt k}$. In our conventions, the vertex operator
$\tilde \VV_{\alpha}=e^{i\sqrt 2 \alpha \tilde \phi}$ has scaling dimension
\eqn\apped{
\Delta(\tilde \VV_{\alpha})=\alpha(\tilde Q+\alpha)
~.}

In the main text, the most important role will be played by 
the physical vertex operators $T_{r,1}$, which we denote as
\eqn\appee{
\TT_r=T_{r,1}=c\bar c ~ \VV_{\frac{r+1}{2\sqrt k}+\frac{\sqrt k}{2}}
\tilde \VV_{\frac{r-1}{2\sqrt k}+\frac{\sqrt k}{2}}
~.}
By definition $\VV_{\alpha}=e^{\sqrt 2 \alpha \phi}$.

\listrefs
\bye